\documentclass{ieeeaccess}

\usepackage{soul}
\usepackage{verbatim}

\ifCLASSINFOpdf
\else
\fi

\usepackage{cite}
\usepackage{soul}
\usepackage{amsmath,amssymb,amsfonts}
\usepackage{algorithmic}
\usepackage{graphicx}
\usepackage{caption}
\usepackage{subcaption}
\usepackage{adjustbox}
\usepackage{tabularx}
\usepackage{breakurl}
\usepackage{longtable}
\usepackage{array}

\usepackage[breaklinks=true]{hyperref}
\usepackage{cite}
\usepackage{textcomp}
\usepackage[table]{xcolor}
\usepackage{enumitem}
\usepackage{multicol}
\usepackage{multirow}

\usepackage{booktabs,tabularx,enumitem,ragged2e}

\usepackage{colortbl}

\def\BibTeX{{\rm B\kern-.05em{\sc i\kern-.025em b}\kern-.08em
    T\kern-.1667em\lower.7ex\hbox{E}\kern-.125emX}}
\begin{document}

\title{Explainable Intrusion Detection Systems (X-IDS): A Survey of Current Methods, Challenges, and Opportunities}

\author{\uppercase{Subash Neupane}\authorrefmark{1},
\uppercase{Jesse Ables\authorrefmark{2}, William Anderson\authorrefmark{3}, Sudip Mittal\authorrefmark{4}}, \uppercase{Shahram Rahimi\authorrefmark{5}}, \uppercase{ Ioana Banicescu\authorrefmark{6}}, and \uppercase{Maria Seale\authorrefmark{7}.}}

\address[1]{Department of Computer Science \& Engineering, Mississippi State University, Mississippi, (email: sn922@msstate.edu)}
\address[2]{Department of Computer Science \& Engineering, Mississippi State University, Mississippi, (email: jha92@msstate.edu)}
\address[3]{Department of Computer Science \& Engineering, Mississippi State University, Mississippi, (email: wha41@msstate.edu)}
\address[4]{Department of Computer Science \& Engineering, Mississippi State University, Mississippi, (email: mittal@cse.msstate.edu)}
\address[5]{Department of Computer Science \& Engineering, Mississippi State University, Mississippi, (email: rahimi@cse.msstate.edu)}
\address[6]{Department of Computer Science \& Engineering, Mississippi State University, Mississippi, (email: ioana@cse.msstate.edu)}
\address[7]{U.S Army Engineer Research and Development Center, Vicksburg, Mississippi, USA, (email: maria.a.seale@erdc.dren.mil)}
\tfootnote{}

\markboth
{Neupane \headeretal}
{Neupane \headeretal}

\corresp{Corresponding author: Subash Neupane (e-mail: sn922@msstate.edu).}

\begin{abstract}
The application of Artificial Intelligence (AI) and Machine Learning (ML) to cybersecurity challenges has gained traction in industry and academia, partially as a result of widespread malware attacks on critical systems such as cloud infrastructures and government institutions. Intrusion Detection Systems (IDS), using some forms of AI, have received widespread adoption due to their ability to handle vast amounts of data with a high prediction accuracy. These systems are hosted in the organizational Cyber Security Operation Center (CSoC) as a defense tool to monitor and detect malicious network flow that would otherwise impact the Confidentiality, Integrity, and Availability (CIA). CSoC analysts rely on these systems to make decisions about the detected threats. However, IDSs designed using Deep Learning (DL) techniques are often treated as black box models and do not provide a justification for their predictions. This creates a barrier for CSoC analysts, as they are unable to improve their decisions based on the model's predictions. One solution to this problem is to design \textit{explainable IDS} (X-IDS).

This survey reviews the state-of-the-art in explainable AI (XAI) for IDS, its current challenges, and discusses how these challenges span to the design of an X-IDS. In particular, we discuss black box and white box approaches comprehensively. We also present the tradeoff between these approaches in terms of their performance and ability to produce explanations. Furthermore, we propose a generic architecture that considers human-in-the-loop which can be used as a guideline when designing an X-IDS. Research recommendations are given from three critical viewpoints: the need to define explainability for IDS, the need to create explanations tailored to various stakeholders, and the need to design metrics to evaluate explanations.
\end{abstract}


\titlepgskip=-15pt

\maketitle

\section{Introduction \& Motivation}

The use of Artificial Intelligence (AI) and Machine Learning (ML) to solve cybersecurity problems has been gaining traction within industry and academia, partly as a response to widespread malware attacks on critical systems, such as cloud infrastructures, government institutions, etc. 
\cite{kwon2019survey, sridhar2011cyber, rajkumar2010cyber, cardenas2009challenges, ahmadian2018information, marino2018adversarial,cardellini2022intrusion,sane2021semantically,mcdole2021deep,wali2021explainable, wang2020explainable}. AI- and ML-assisted cybersecurity offers data-driven automation that could enable security systems to identify and respond to cyber threats in real time. Many of these AI-based cyber defense systems are hosted in an organizational Cyber Security Operations Center (CSoC).
CSoCs operated by security analysts act as a cybersecurity information hub and a defense base. Here the task is to orchestrate different security systems that are a part of an organization's overall cybersecurity framework, many of which have AI components. Examples of these security systems include Security Information and Event Management (SIEM) systems, vulnerability assessment solutions, governance, risk and compliance systems, application and database scanners, Intrusion Detection Systems (IDS), user and entity behavior analytics, Endpoint Detection and Remediation (EDR), etc. Here the security analysts maintain an 
``organizational state”, keeping themselves one step ahead of the attackers to prevent potential intrusions \cite{raytheon_2017}. 


The term ``intrusion detection” originated in the early 1980s with James Anderson's seminal paper \cite{anderson1980computer}. Dorothy E. Denning \cite{denning1987intrusion}, following Anderson's work, proposed the first functional IDS in the mid-1980s. An IDS is a software or hardware security system that automates the process of monitoring and analyzing events occurring within a computer system or network for indications of potential security problems before they inflict widespread damage \cite{bace2001intrusion, wu2010use}. 

In general, an intrusion results in a breach of at least one of the principles: \textit{Confidentiality}, \textit{Integrity}, or \textit{Availability} (CIA). These tenets of security are used when protecting modern data infrastructure. They refer to the permissions to access or modify data, the prevention of improper data modification, and the ability to access data. The objective of an IDS is to detect misuse, unauthorized use (outsider without authorization), and abuse (abusing privilege-e.g., insider threat) within an organization, and much research has been done to improve the operational capacity of these IDS \cite{mukherjee1994network,lee1999data, buczak2015survey}.

The literature shows that numerous IDSs have been developed through the application of a variety of techniques from an array of disciplines, including statistical methods, ML techniques, and others \cite{belouch2018performance}. At present, ML and Deep Learning (DL) techniques are widely used to develop IDS because of their ability to attain a high detection rate \cite{wu2010use}. This adoption is also attributed to the fact that IDSs based on ML/DL techniques are much more efficient, accurate, and extendable as compared to their counterparts developed using other techniques.
The surveys in \cite{aminanto2016deep, kim2017deep, kwon2019survey} primarily focus on intrusion detection techniques based on deep learning. However, \textit{the techniques described in the preceding surveys are deficient in their ability to explain their inference processes and final results}, and they are \textit{frequently treated as a black box by both developers and users} \cite{xu2019explainable}. As a result, there is growing concern about the possibility of bias in these models, which necessitates the requirements for model transparency and post-hoc explainability \cite{gunning2019darpa}. Unfortunately, the majority of black box IDS described in the literature are opaque and much is needed to augment transparency.

It is apparent that these opaque/non-transparent models can achieve \textit{impressive prediction accuracies}; however, they lack justification for their predictions.   This is due to their nested and non-linear structure, which makes it difficult to identify the precise information in the data that influences their decision-making \cite{samek2017explainable}. Such a \textit{lack of understanding about the inner workings of opaque AI models} or an inability to traverse back from the outputs to the original data raises user trust issues \cite{marshan2021artificial}. This black box nature of the models creates problems for several domains in which AI or components of AI are integrated \cite{berk2013statistical}. 
For example, in the context of an IDS, CSoCs analysts are tasked with the responsibility of analyzing IDS alerts for a variety of purposes, including alert escalation, threat and attack mitigation, intelligence gathering, and forensic analysis among others \cite{Nguyen2019GEEAG}. The lack of explanation of alerts generated by an IDS creates a barrier for task analysis and subsequently impedes decision-making.


In addition to the issues of \textit{transparency} and \textit{trust} surrounding  AI systems, there exists yet another issue referred to as the problem of \textit{decomposability}, specifically for systems built with DL models (See Section \ref{subsubsection:decomposition_based} for IDS based on the decomposition approach). AI systems that are designed using DL techniques are difficult to interpret due to their inability to be decomposed into intuitive components \cite{lipton2018mythos}. The difficulty in interpreting DL models jeopardizes their actual use in production, as the computation behind their decisions are unknown \cite{wang2020explainable}. \textit{Explainable AI} (XAI) seeks to remedy this and other problems.

According to the Defense Advanced Research Projects Agency (DARPA), XAI systems are able to explain their reasoning to a human user, characterize their strengths and weaknesses, and convey a sense of their future behavior \cite{gunning2019darpa}. In this sense, by justifying specific decisions, XAI systems aid users in comprehending the model and assisting them in maintaining and effectively using it. On the other hand, transparency about predictions contributes to the development of trust in a system's intended behavior and provides users with confidence that they are performing tasks correctly.

Transparency is an open problem in the field of intrusion detection. Cybersecurity professionals now frequently make decisions based on the recommendations of an AI-enabled IDS. Therefore, the predictions made by the model should be understandable \cite{wang2020explainable}. For instance, when an IDS model is presented with zero-day attacks, the model may misclassify the attacks as normal, resulting in a system breach. Understanding why specific samples are misclassified is the first step toward debugging and diagnosing the system. It is critical to provide detailed explanations for such misclassifications, so as to determine the appropriate course of action to prevent future attacks \cite{marino2018adversarial}. 
Therefore, an IDS should go beyond merely detecting intrusions-i.e., it should provide reasoning for the detected threat. The explanations in the form of correlations of various factors (for example, time of intrusion, type, suspicious network flow) influencing the predicted outcome can assist cybersecurity analysts in quickly analyzing tasks and making decisions \cite{Nguyen2019GEEAG}.

The goal of \textit{XAI in the field of intrusion detection} is to build operator trust and allow for more control of autonomous AI subsystems. Explainable Intrusion Detection Systems (X-IDS) 
will help build trust in these systems while also aiding CSoC analysts in their task of defending systems.

The major contributions of the paper are as follows:
\begin{itemize}
    \item We present the state-of-the-art of the XAI approach and discuss the critical issues that surround it, most importantly, how these issues relate to the intrusion detection domain. We propose a taxonomy based on a literature review to help lay the groundwork for formally defining explainability in intrusion detection.
    \item A comprehensive survey of the current landscape of X-IDS implementations is presented, with an emphasis on two major approaches: black box and white box. The distinction between the two approaches is discussed in detail, as is the rationale for why the black box approach with post-hoc explainability is more appropriate for intrusion detection tasks.
    \item We propose a generic explainable architecture with a user-centric approach for designing X-IDS that can accommodate a wide variety of scenarios and applications without adhering to a specific specification or technological solution.
    \item We discuss the challenges inherent in designing X-IDS and make research recommendations aimed at effectively mitigating these challenges for future researchers interested in developing X-IDS.
\end{itemize}

The remainder of this paper is organized as follows. Section \ref{xai} provides the background on explainable artificial intelligence (XAI). Section \ref{overview} summarizes our survey and taxonomy. Following that, in Section \ref{xaiids}, we review the literature on black box and white box X-IDS approaches. Section \ref{design_xai_ids} introduces a generic X-IDS architecture that future researchers can use as a guide. Section \ref{recommendation} identifies research challenges and makes recommendations to future researchers. Finally, Section \ref{conclusion} concludes this survey.

\section{Explainable Artificial Intelligence (XAI)}
\label{xai}

The definition of what constitutes an explanation in AI remains an open research question. In the available literature, there are various definitions of the `explainable AI' (XAI). Lent et al. \cite{van2004explainable} defines XAI as a system that \textit{``provides an easily understandable chain of reasoning from the user's order, through the system's knowledge and inference, to the resulting behavior''}. The authors of this work used the term XAI to describe their system's ability to explain the behavior of AI-controlled entities. However, the recent definition by DARPA \cite{darpa2016broad} indicates XAI as the intersection of different areas including machine learning, human computer interface, and end user explanation.


The problem of explainability dates back to the mid-1970s. Even at the time, Moore and Swartout \cite{moore1988explanation} argued that explanation capabilities are not only desirable but necessary for expert systems to succeed. 
However, the focus in the scientific and research community shifted towards the development of AI models that could produce exceptional predictions, while little emphasis was given to addressing the reason these models were able to produce such results. Today, XAI has gained renewed attention from the research community and application users. The renaissance of the XAI research is attributed to the integration of AI with machine learning (ML) across industries and its impact on the critical decision-making process, despite its inability to provide information about the reasoning behind specific decisions \cite{adadi2018peeking}.  

XAI has the potential to offer significant benefits to a broad range of domains that rely on artificial intelligence systems. Currently, XAI is being used in mission-critical systems and defense \cite{darpa2016broad, van2004explainable}. To foster the trust of AI systems in the transportation domain, researchers are proposing explanations systems \cite{haspiel2018explanations}. Some works based on image processing with explainability is found in \cite{lorente2021explaining, li2020deep, martinez2020interpretable, ponn2020identification}. Transparency regarding decision-making processes is critical in the criminal justice system \cite{deeks2019judicial, berk2013statistical}. Various explainable methods for judicial decision support systems have been proposed by authors in \cite{loyola2019understanding, zhong2019explainable, vlek2016method}. Model explainability is essential for gaining trust and acceptance of AI systems in high-stakes areas, such as healthcare, where reliability and safety are critical \cite{holzinger2017we, gade2019explainable}. Medical anomaly
detection \cite{itani2020one}, healthcare risk prediction system \cite{lindsay2020explainable, pintelas2020explainable, prifti2020interpretable, lundberg2017explainable},
genetics \cite{huang2020quantitative, anguita2020explainable}, and healthcare image processing \cite{muddamsetty2021expert, graziani2020concept, rio2020understanding} are some of the areas that are moving towards adoption of XAI. Another area is finance, such as AI-based credit score decisions \cite{adadi2018peeking, chun2021study} and counterfeit banknotes detection \cite{han2019joint}. 
Support for XAI in academia 
for evaluation tasks are found in \cite{amparore2021trust, van2021evaluating, sokol2020explainability}. Lastly, in the entertainment industry XAI for recommender systems is found in the works of \cite{rutkowski2019explainable, wang2019explainable, zhao2019personalized}.

Arrieta et al. \cite{arrieta2020explainable} argue that one of the issues that hinders the
establishment of common ground for the meaning of the  term `explainability' in the context of AI is the interchangeable misuse
of `interpretability’ and `explainability’ in the literature. Interpretability is the ability to explain or convey meaning in human-comprehensible terms \cite{arrieta2020explainable}. This translates into the ability of a human to understand the model’s reasoning without the need for additional explanations \cite{guidotti2018survey}. On the other hand, explainability is associated with the concept of explanation as a means of interface between humans and a decision-maker (model) that is both accurate and comprehensible to humans \cite{guidotti2018survey}. In this sense, if system users need an explanation as a proxy system to understand the reasoning process, that explanation is precisely represented by the XAI. 

A central concept that emerges from all the preceding
definitions of the XAI is `understandability’, which is the degree to
which a human can comprehend a decision made with respect
to a model. However, understandability is tightly coupled with the characteristics of the system's users. For instance, whether or not the explanation made the concept clear or simple to understand is entirely dependent on the audience. 

Despite the widespread recognition of the importance of explainability, researchers are struggling to establish universal, objective criteria for developing and validating explanations \cite{miller2019explanation}. This is because XAI is plagued by inherent challenges that need addressing to foster its development. These include (\romannumeral 1)  achieving consensus on the right notion of model explainability, (\romannumeral 2) identifying and formalizing explainability tasks from the perspectives of various stakeholders, and (\romannumeral 3) designing measures for evaluating explainability techniques \cite{gade2020explainable}.



\begin{figure}[htbp]
    \centering
    \includegraphics[scale=0.50]{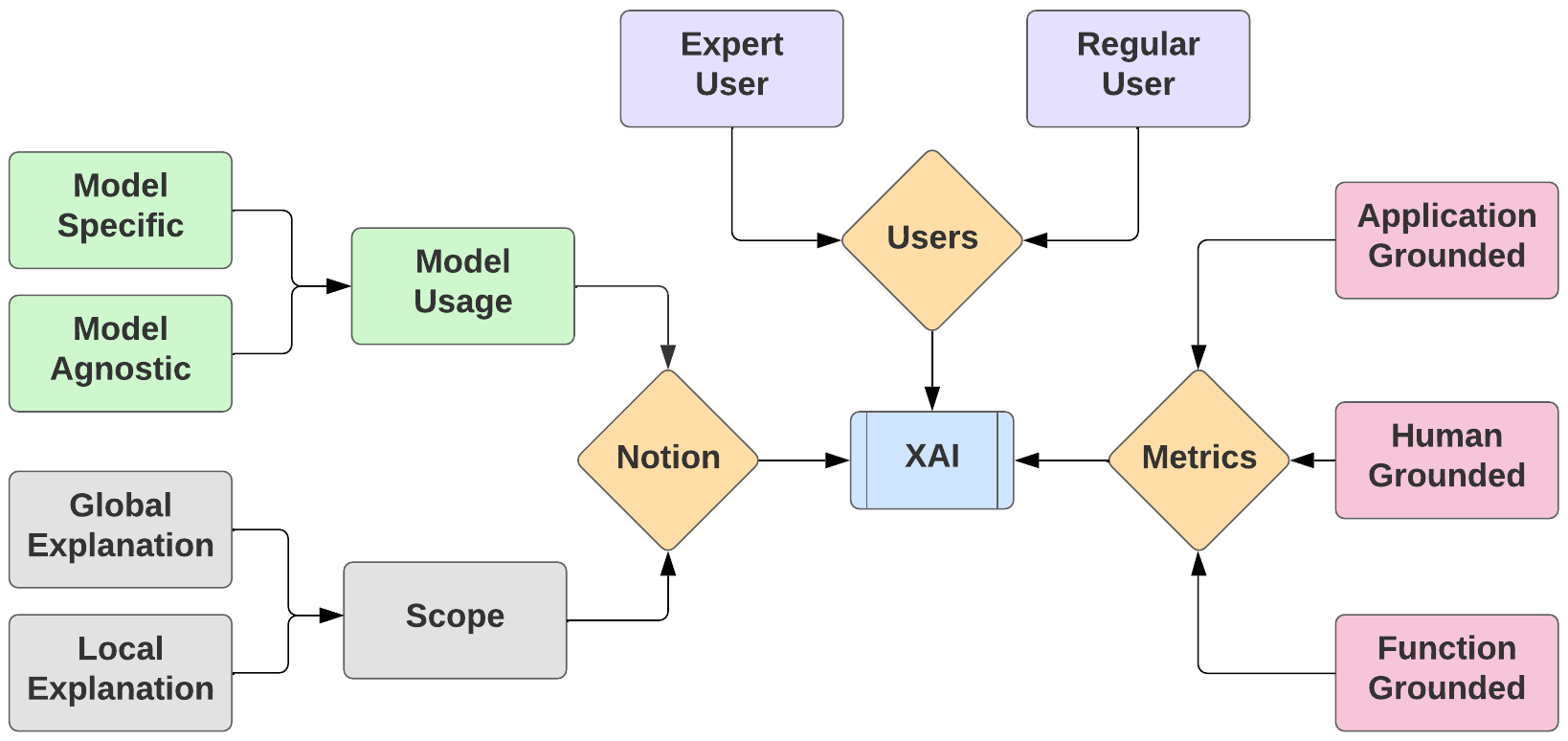}
    \caption{A taxonomic approach to explainability definition based on explainability concepts, formalizing explainability tasks from the standpoint of stakeholders, and evaluating explainability techniques. Green represents model dependency, while grey represents the scope of the explanations. Light purple represents various types of stakeholders in the IDS ecosystem. Pink represents techniques to evaluate explanations.}
        \label{fig: explainablility_taxonomy}
\end{figure}

To address these challenges, we propose a taxonomy as depicted in Figure \ref{fig: explainablility_taxonomy} based on our literature review to lay the groundwork for formally developing and validating explanations. In the following subsections, we describe in detail the concepts related to XAI including its notions, its meaning to various stakeholders of the system, and metrics to evaluate explanations.

\subsection{Notions of Explainability} Several approaches to explanation methods have been proposed by different authors in the pursuit of explaining AI systems. The authors in \cite{guidotti2018survey} conducted a survey of black box specific explainability methods and proposed a taxonomy for XAI systems based on four characteristics: (\romannumeral 1) the nature of the problem; (\romannumeral 2) the type of explainer used; (\romannumeral 3) the type of black box model processed by the explainer; and (\romannumeral 4) the type of data supported by the black box. \\
\indent In another work \cite{vilone2021notions}, the authors presented notions related to the concept of explainability in two clusters. 
The first cluster refers to attributes of explainability – it contains criteria and characteristics used by scholars in trying to define the construct of explainability. The second cluster refers to the theoretical approaches for structuring explanations. Das and Rad in \cite{das2020opportunities} proposed a taxonomy for categorizing XAI techniques based on explanation scope, algorithm methodology, and usage.
Similarly, the authors in \cite{adadi2018peeking} surveyed over 180 articles related to explainability and categorized explainability using three criteria: the complexity of interpretability, the scoop of interpretability, and model dependency. The first criterion emphasizes the difficulty of interpreting and explaining complex models, such as those based on deep learning. The second criterion differentiates between local and global explanations, while the third criterion discusses model-specific and model-agnostic explanations. On the other hand, Pantelis et al. in \cite{linardatos2021explainable} divided explainability methods into four groups based on: the data types used, the scope of explanation, the purpose of explanation, and the model usage. 

A common category found in the literature regarding the taxonomy of explainability is the \textit{scope of explainability} and \textit{model dependency}. The following subsections describe these categories in greater detail.
\subsubsection{Local explainability}\label{localxai} 
The ability to explain a single prediction or decision is an example of local explainability. This explainability is used to generate a unique explanation or justification of the specific decision made by the model \cite{adadi2018peeking}. Some of the local explanation methods include the Local Interpretable Model Agnostic Explanation (LIME) \cite{ribeiro2016should}, the Anchors \cite{ribeiro2018anchors} and the Leave One Covariate Out (LOCO) \cite{lei2018distribution}. The LIME was originally proposed by Ribeiro et al. \cite{ribeiro2016should}.  The authors of this paper used a surrogate model to approximate the predictions of the black box model.  Rather than training a global surrogate model, the LIME uses a local surrogate model to interpret individual predictions.
The explanation produced by LIME is obtained by using Equation \ref{eq:local_surrogate}. 

\begin{equation} \label{eq:local_surrogate}
\xi(x)=\underset{g \in G}{\operatorname{argmin}}\left\{\mathcal{L}\left(f, g, w^{x}\right)+\Omega(g)\right\}
\end{equation}

where:
\begin{quote}
 ${g}$ represents the model of explanation for the occurrence ${x}$, (e.g., linear regression); ${G}$ denotes a class of potentially interpretable models, such as linear models, decision tree; ${\mathcal{L}}$ denotes the loss function (e.g., mean squared error), which is used to determine how close the explanation model's predictions are to the original model's predictions; ${f}$ denotes the original model; $w^{x}$ specifies the weighting factor between the sampled and original data;
 ${\Omega(g)}$ captures the complexity of model ${g}$.
\end{quote}

To explain the behavior of complex models with high precision rules called \textit{Anchors}, representing local, \textit{sufficient conditions} for predictions, the same authors proposed an extension to the LIME method in \cite{ribeiro2018anchors}. Another popular technique for generating local explanation models with local variable importance measures is LOCO \cite{lei2018distribution}.

Lundberg et al. \cite{lundberg2017unified} proposed a game-theoretic optimal solution based on Shapley values for model explainability referred to as Shapely Additive Explanations (SHAP). SHAP calculates the significance of each feature in each prediction. The authors have demonstrated the equivalence of this model among various local interpretable models including LIME \cite{ribeiro2016should}, Deep Learning Important FeaTures (DeepLIFT) \cite{shrikumar2017learning}, and LayerWise Relevance Propagation (LRP) \cite{binder2016layer}. The SHAP value can be computed for any model, not just simple linear models. SHAP specifies the explanation for an instance \textit{x} as seen in Equation \ref{eq:shap}:

\begin{equation}\label{eq:shap}
g\left(z^{\prime}\right)=\phi_{0}+\sum_{j=1}^{M} \phi_{j} z_{j}^{\prime}
\end{equation}

where:
\begin{quote}
 ${g}$  represents explanation model; $z^{\prime}$  is the coalition vector of the simplified features, and $z^{\prime} \in\{0,1\}^{M}$ (The 1 in $z^{\prime}$ indicates that the new data contains identical features to the original data, while the 0 indicates that the new data contains features that are distinct from the original data);  ${M}$  denotes the maximum size of a coalition;  $\phi_{j} \in \mathbb{R}$ denotes the attribute for the feature j in the instance ${x}$. This is known as the Shapley value. If $\phi_{j}$ is a large positive number, it indicates that feature ${j}$  has a significant positive effect on the model's prediction. $\phi_{0}$ represents the model output with all simplified inputs toggled off (i.e. missing).
\end{quote}

\subsubsection{Global explainability} The global explainability of a model makes it easier to follow the reasoning behind all the possible outcomes. These models shed light on the model's decision-making process as a whole, resulting in an understanding of the attributions for a variety of input data \cite{das2020opportunities}.

The LIME \cite{ribeiro2016should} model was extended with a `submodular pick algorithm' (SP-LIME) in order to comprehend the model's global correlations. By providing a non-redundant global decision boundary for the machine learning model, LIME provides a global understanding of the model from individual data instances using a submodular pick algorithm. 

Concept Activation Vectors (CAVs) proposed by Kim et al. \cite{kim2018interpretability} is another global explainability method. This model can interpret the internal states of a neural network in the human-friendly concept domain. In 
another work, Yang et al. \cite{yang2018global}  proposed a novel method, the Global Interpretation via Recursive Partitioning (GIRP), to construct a global interpretation tree based on local explanations for a variety of machine learning models. Other methods of global explanation include an explanation by information extraction \cite{valenzuela2018lightly}. In this study, the authors propose a method of information extraction that is only lightly supervised and provides a global interpretation. They demonstrated that interpretable models can be generated when representation learning is combined with traditional pattern-based bootstrapping.

\subsubsection{Model-specific interpretability} The use of model-specific interpretability methods is restricted to a limited number of model classes. With these methods, we are restricted to using only models that provide a specific type of interpretation, which can reduce our options for using more accurate and representative models. 

\subsubsection{Model-agnostic interpretability}\label{subsection:model_agnostic}

Methods that are model agnostic are not tied to any specific type of ML model, and are by definition modular, in the sense that the explanatory module is unrelated to the model for which it generates explanations. Model-agnostic interpretations are used to interpret artificial neural networks (ANNs) and can be local or global. In their survey \cite{das2020opportunities}, the authors argue that a significant amount of research in XAI is concentrated on model-agnostic post-hoc explainability algorithms, due to their ease of integration and breadth of application. Based on other reviewed papers, the authors \cite{adadi2018peeking} broadly categorize the techniques of model-agnostic interpretability into four types, including visualization, knowledge extraction, influence methods, and example-based explanations. 

\subsection{Formalizing explainability tasks from the user perspectives}
To be explainable, a machine learning model must be human-comprehensible. This presents a challenge for the development of XAI because it entails communicating a complex computational process to humans. The interpretable element that serves as the foundation of explanation is highly dependent on the question of \textit{``who”} will receive the explanation. The authors in \cite{rosenfeld2019explainability} identified three targets of explanation, including regular user, expert user, and the external entity. According to the authors, an explanation should be specific to user types. For instance, in a legal scenario, the explanation must be made to the expert users, not the regular users. On the other hand, if explanations are geared towards regular users, then the chance of developing trust and acceptance of XAI methods is high.\\ 
\indent Keeping humans in the loop determines the overall explainability value \cite{adadi2018peeking}. The authors of this study emphasize the significance of humans-in-the-loop approach for explainable systems from two perspectives: Human-like explanation and Human-friendly explanation. The first aspect focuses on how to produce explanations that simulate the human cognitive process, while the second aspect is concerned with developing explanations that are centered on humans.

Section \ref{design_xai_ids} discusses the importance of human-centered design when developing X-IDS systems, and Section \ref{recommendation_stakeholder} examines the explainability requirements imposed by various stakeholders in the IDS ecosystem. 

\subsection{Measures for evaluating explainability techniques} \label{exp_metric}
There have been few studies on evaluating explanations and quantifying their relevance despite the growing body of research that produces explainable ML methods. 
Doshi-Velez and Kim \cite{doshi2017towards} proposed the three classes as evaluation methods for interpretability, including application-grounded, human-grounded, and functionally grounded methods. Application-grounded evaluation is concerned with the impact of the interpretation process's results on the human, domain expert, or end user, in terms of a well-defined task or application. Human-grounded evaluation is concerned with conducting simplified application-grounded evaluation where experiments are run with regular users rather than domain experts. Functionally grounded evaluation does not require human subjects, and rather uses formal, well-defined mathematical definitions of interpretability to determine the method's quality.

On the other hand, in \cite{gilpin2018explaining}, the authors outline three different evaluation criteria of explanations for deep networks, such as processing, representation, and explanation producing. The first criterion includes techniques that simulate data processing to generate insights about the relationships between a model's inputs and outputs. The second criterion describes an approach on how data is represented in networks and explains the representation. The third criterion states that the explanation-producing systems can be evaluated according to how well they match user expectations. Other types of evaluation criteria found in the body of literature include completeness compared to the original model, ability to detect models with biases, completeness as measured on a substitute task, and human evaluation.

Another significant piece of work that could serve as a benchmark for evaluating explanations is the Florida Institute for Human and Machine Cognition's psychological model of explanation (IHMC)\cite{gunning2019darpa}. Section \ref{recommendation_evaluation_metrics} provides greater detail about this proposed model.

Next, we describe our survey approach and develop a taxonomy for X-IDS grounded in the current literature.

    \section{Survey \& Taxonomy}\label{overview}

The term \textit{intrusion} refers to any unauthorized activity occurring within a network or system. An IDS is a collection of tools, methods, and resources that assist CSoC analysts in identifying, assessing, and reporting intrusions. Intrusion detection is typically a component of protection that surrounds a system and is not a stand-alone protection measure \cite{butun2013survey}. IDS are classified according to where they look for intrusive behavior: \textit{host-based} or \textit{network-based}. A host-based IDS monitors traffic that is originating from and coming to a specific host. Network-based IDS are strategically positioned in a network to analyze incoming and outgoing communication between network nodes. 

IDS are categorized based on three detection techniques: \textit{signature-based}, \textit{anomaly-based}, and \textit{hybrid}. Signature-based IDS monitors network traffic and compares it to a database of known malicious threats' signatures or attributes. However, they are incapable of detecting zero-day attacks, metamorphic threats, or polymorphic threats \cite{sharma2014evolution}. On the other hand, anomaly-based IDS look for patterns in data that do not conform to expected behavior \cite{chandola2009anomaly}, allowing them to recognize such threats. However, these detection systems are susceptible to higher false positive rates because they may categorize previously unseen, yet legitimate, system behaviors as anomalies \cite{buczak2015survey}. Hybrid IDS integrate both signature-based and anomaly-based detection methods, which allows for an increased detection rate of known intrusions, the ability to detect unseen intrusions, and reduce false positives.

\begin{figure*}[h!]
    \centerline{\includegraphics[scale=0.68]{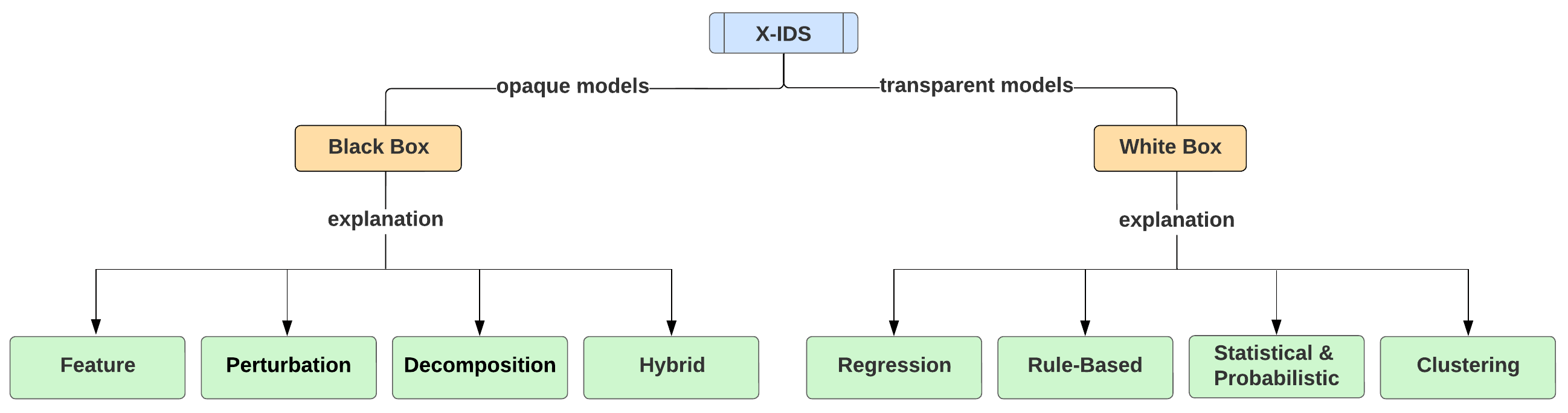}}
    \caption{An overview of our proposed taxonomy. We categorize X-IDS techniques into two families, white box and black box. White box approaches encompass the techniques of Regression, Rule-Based, Clustering, and Statistical \& Probabilistic Methods. Black box approaches encompass Feature, Perturbation, Decomposition, and Hybrid approaches. These approaches define the method of explainability to interpret the model's decision process.}
    \label{tax_fig}
\end{figure*}

As previously stated, prior work has focused on XAI from the lens of explainability, qualifying the definitions of  \textit{notion}, \textit{users}, and \textit{metrics} (See Section \ref{xai}). This survey follows that direction by creating a taxonomy surrounding current XAI techniques for IDS. The focus is on their relevance and applicability to the domain of intrusion detection, with a particular emphasis on the current hierarchy of families, strengths and weaknesses, and any challenges or assumptions that come with their application. A summary of our taxonomy can be seen in Figure \ref{tax_fig}. The two primary families of XAI techniques are those of white box models and black box models which greatly affect our survey taxonomy for approaches to X-IDS. The survey of existing systems based on the taxonomy in Figure \ref{tax_fig}, is available in Section \ref{xaiids}. Next, we describe the salient features of white box models and black box models.

\subsection{Salient Features of White Box Techniques} \label{overview_whitebox_approach}
\textit{White box models} provide results that are easy to understand \cite{loyola2019black}. This \textit{easy to understand} condition is typically defined as an explainable outcome understood by an expert in the field. In practice, this definition is more associated with the popular suite of machine learning models that existed prior to the rise in popularity of neural network based approaches. White box models, while generally not as efficacious as their black box counterparts, bring a layer of transparency that is intrinsic to their decision process. This trait is often preferred, if not a requirement, in domains where the decision system is sensitive or requires a high degree of auditing. These models cover a wide variety of techniques that fall into four distinct families: \textit{Regression}, \textit{Rule-Based}, \textit{Clustering}, and \textit{Statistical \& Probabilistic Methods}.

Regression-based approaches comprise the family of regression analysis. These approaches have a well formed background of statistical support and maturity. Therefore, these models are most often employed in the early stages of modeling, in the pipelines of more complex models, and in domains where scrutiny and transparency are of paramount importance. Although not a focal point of comparison for this paper, regression models are highly computationally efficient, allowing for rapid construction, as well as deployment into low-resource systems where detection time is critical, such as IoT edge devices. Regression approaches can be split into Parametric Regression and Non-parametric Regression. The former enforces a constraint on model expectation via a restriction on the parameters of the model, making this modeling approach best for when certain assumptions can be met. The latter enforces no such constraint, which decreases overall interpretability but increases the application to a wider variety of data and assumptions. Popular regression techniques are Linear Regression (LR), Logistic Regression (LoR), various non-linear models, Poisson Regression, Kernel Regression (KR), and Spline Smoothing.

Rule-based approaches leverage a learned set of rules as a means of the model decision process, and thusly, model explainability. Rule based explanations are perhaps the most practical, as they mimic the human decision making process when it comes to defining an anomaly. This process also allows learned rules to then be incorporated into Signature-Based IDS (SIDS), allowing Anomaly-Based IDS (AIDS) to serve as zero-day identifiers and rule miners. Rule-based approaches benefit from the allowance of a very tight definition of rules, known as hard rules or crisp rules, or for a relaxed fuzzy-rule based approach, allowing flexibility and further statistical inference to be rendered on them. A popular approach to modeling for rule-based explanations is the Decision Tree and its many variants. 

Statistical \& Probabilistic Methods is a broad category for the numerous statistical models of reasoning that exist in the literature. Notably, many of these methods have seen a decline in use as a compliment to the rise in popularity of various black box methods. These less frequently used methods are appropriate for application in specific scenarios or in larger pipelines for multi-stage IDS. Examples of such approaches include moment-based approaches, statistical ensembles, Markov Models, Baysian Networks, and others which are covered more specifically in \ref{Statistical and Probabilistic Models}.

Clustering-based approaches use supervised or unsupervised learning to aggregate similar data objects. This \textit{similar} condition is defined by a similarity, or dissimilarity, measure. Traditionally these methods are defined on distance based metrics such as Euclidean, Manhattan, Cosine Measure, Pearson coefficient, and many others. Other attempts to define similarity have had success in the graph representation domain, using graph-based clustering algorithms to accomplish this task. Clustering, due to its ability to be leveraged as an unsupervised learner, still retains a high degree of use due to the importance of data mining for intrusion detection. Examples of popular clustering algorithms are K-Means, Self-Organzing Maps (SOMs), Density-Based Spatial Clustering of Applications with Noise (DBSCAN), Agglomerative Clustering, and Spectral Clustering.

\subsection{Salient Features of Black Box Techniques}

\textit{Black box models} are models where the decision systems are considered opaque \cite{guidotti2018survey}. These systems, composing nearly all of the state-of-the-art, are limited due to the lacking ability of model inspection and evaluation. Therefore, if these systems are to be utilized in decision sensitive domains, i.e, those whose applications require safety, privacy, and fairness, some degree of exploration and evaluation of their decision process must be possible. Currently, there exists no singular solution to the black box inspection problem. However, many candidate explanations have emerged, exploring and exploiting various aspects of the machine learning process to create explanations for black box models. These candidate explanations currently fall into four distinct families:
\textit{Feature}, \textit{Perturbation}, \textit{Decomposition}, and \textit{Hybrid}.

Feature-based explanations target features as the method of explanation. The goal of feature attribution is to determine how much each feature is responsible for the output prediction. Features were one of the first methods of explainability in black box models due to their impact on model performance and human interpretable relevance. Examples of popular feature based explanations are Partial Dependence Plot (PDP)\cite{friedman2001greedy}, Accumulated Local Effects (ALE)\cite{apley2020visualizing}, Individual Conditional Expectations (ICE)\cite{goldstein2015peeking}, H-statistic\cite{friedman2008predictive}, and SHapley Additive exPlanations (SHAP)\cite{lundberg2017unified}.

Perturbation-based explanations study changes to the output space with perturbations to the input space. Due to this property, perturbation techniques can be deployed to any general input space, such as tabular data, images, or text. In particular, model sensitivity to feature perturbations has long been regarded as a measure of feature importance. Saliency maps, Randomized Input Sampling for Explanation of Black Box Models (RISE)\cite{petsiuk2018rise}, and Local Interpretable Model-Agnostic Explanations (LIME)\cite{ribeiro2016should} are popular perturbation methods.

Decomposition-based explanations decompose the original model prediction. Much like the previous two methods, the goal of decomposition is to allocate a measure of importance to the input space; however, this method does so by the decomposition of a model signal, such as the model's gradients. This is predicated on the assumption that large gradients play a role in shaping explanations. However, gradients are not the only method of decomposition. Many approaches exist, such as Gradient * Input \cite{shrikumar2016not}, Integrated Gradients (IG) \cite{sundararajan2017axiomatic}, Grad-CAM \cite{selvaraju2017grad}, DeepLIFT \cite{shrikumar2017learning}, Deep Taylor Decomposition (DTD) \cite{montavon2017explaining} and Layerwise Relevance Propagation (LRP)\cite{binder2016layer} . 

Hybrid-based explanations encapsulate a type of model construction often demonstrated in pipelined machine learning architectures. These models can range from ensembles, a blend of white box and black box approaches working in tandem, to carefully composed IDS pipelines encapsulating many of the best state-of-the-art approaches. Therefore, hybrid approaches present the most variability of explanations, with respect to methodology, location of explanations, and application. 

Next, we use the taxonomy showcased in Figure \ref{tax_fig}, to present a literature survey on approaches to X-IDS.

\section{Approaches to Explainable IDS (X-IDS)}\label{xaiids}

As per the survey overview presented in Section \ref{overview}, and the taxonomy showcased in Figure \ref{tax_fig}, we will now describe in detail the black box and the white box approaches to XAI in intrusion detection systems.

\subsection{Black Box X-IDS Models} 

Guidotti et al. \cite{guidotti2018survey}, describes a black box predictor as ``a data-mining and machine learning obscure model, whose internals are either unknown to the observer or are known but are uninterpretable by humans''. A black box model is not explainable by itself. Therefore, to make a black box model explainable, we have to adopt several techniques to extract explanations from the inner logic or the outputs of the model. 

To survey the IDS landscape with respect to explainability, we have further divided the literature into different categories of XAI black box models: feature based, perturbations based, decomposition based, and hybrid approaches. These classifications are based upon how explanations are generated. 
A detailed literature overview is also available in Table \ref{table:blackbox_table}. 

\begin{table*}
\centering
{
   \renewcommand{\arraystretch}{1.30}%
    \begin{tabularx}{\textwidth}{X|X|X|X}
    \hline
    \rowcolor{lightgray!20!}
    \textbf{Paper Title} &
    \textbf{Focus/Objective} &
    \textbf{Contribution} &
    \textbf{Limitation} \\
    \hline
    \rowcolor{cyan!20!}
    \multicolumn{4}{c}{Feature based IDS} \tabularnewline
    \hline
    An Explainable Machine Learning Framework for Intrusion Detection Systems \cite{wang2020explainable} &
    Locally and globally explainable NN using SHAP for IDS. &
    \begin{minipage}[t]{\linewidth}
    \begin{itemize}[leftmargin=*]
      \item Framework that creates both local and global explanations.
      \item First use of SHAP in the field of IDS.
      \item Comparison between one-vs-all classifier and multi-class classifier.
    \end{itemize}
    \vspace{1mm}
    \end{minipage} &
    \begin{minipage}[t]{\linewidth}
    \begin{itemize}[leftmargin=*]

        \item More intrusion detection datasets should be tested.
        \item SHAP cannot work in real-time.
        \item SHAP needs to be tested on more robust attacks.
    \end{itemize}
    \vspace{1mm}
    \end{minipage}
     \\ 
    \hline

    Domain Knowledge Aided Explainable Artificial Intelligence for Intrusion Detection and Response \cite{islam2019domain} &

    Use CIA principles on data to improve both generalizability and explainability of a model &
    \begin{minipage}[t]{\linewidth}
    \begin{itemize}[leftmargin=*]

      \item Method for the collection and use of Domain Knowledge in IDS
      \item Use CIA principles to aid in explainability
      \item Domain Knowledge increase generalizability
    \end{itemize}
    \vspace{1mm}
    \end{minipage} &
    \begin{minipage}[t]{\linewidth}
    \begin{itemize}[leftmargin=*]

        \item Domain Knowledge is applied to a specific dataset. New mappings may be needed on new datasets.
        \item More datasets need to be tested.
    \end{itemize}
    \vspace{1mm}
    \end{minipage}
     \\
    \hline

     An Explainable Machine Learning-based Network Intrusion Detection System for Enabling Generalisability in Securing IoT Networks\cite{Sarhan2021AnEM} &

    Explore explainability in IDS by comparing two different IDS feat.  &
    \begin{minipage}[t]{\linewidth}
    \begin{itemize}[leftmargin=*]

      \item Evaluate twe different Network Intrusion Detection datasets: NetFLow and CICFlowMeter.
      \item Creation of two new datasets in the CICFlowMeter format.
      \item An explainable analysis is performed using SHAP.
    \end{itemize}
    \vspace{1mm}
    \end{minipage} &
    \begin{minipage}[t]{\linewidth}
    \begin{itemize}[leftmargin=*]

        \item Explanations are only done using SHAP.
        \item No analysis on the performance of the explainer.
    \end{itemize}
    \vspace{1mm}
    \end{minipage}
     \\
    \hline

    Explaining Anomalies Detected by Autoencoders Using SHAP\cite{Antwarg2019ExplainingAD} &

    Use SHAP to create custom explanations for anomalies found with an autoencoder. &
    \begin{minipage}[t]{\linewidth}
    \begin{itemize}[leftmargin=*]

      \item Method for explaining anomalies found by an autoencoder.
      \item Preliminary experiment with real word data and domain experts.
      \item Suggest methods for evaluating explanations.
    \end{itemize}
    \vspace{1mm}
    \end{minipage} &
    \begin{minipage}[t]{\linewidth}
    \begin{itemize}[leftmargin=*]

        \item Custom explanation lacks any form of visualization to aid the user.
    \end{itemize}
    \vspace{1mm}
    \end{minipage}
     \\
    \hline
    \rowcolor{cyan!20!}
    \multicolumn{4}{c}{Perturbation based IDS} \tabularnewline
    \hline

    A New Explainable Deep Learning Framework for Cyber Threat Discovery in Industrial IoT Networks \cite{khan2021new} &

    Explainable Intrusion Detection in the field of IoT &
    \begin{minipage}[t]{\linewidth}
    \begin{itemize}[leftmargin=*]

      \item Conv-LSTM-based autoencoder for time-series attacks
      \item Detects zero-day attacks
      \item Sliding window technique that increases accuracy of CNN and LSTM model XAI concepts to improve trust
    \end{itemize}
    \vspace{1mm}
    \end{minipage} &
    \begin{minipage}[t]{\linewidth}
    \begin{itemize}[leftmargin=*]

        \item Tested only on a single dataset
        \item Considers only univariate time-series data
    \end{itemize}
    \vspace{1mm}
    \end{minipage}
     \\
    \hline

    An Adversarial Approach for Explainable AI in Intrusion Detection Systems \cite{marino2018adversarial} &

    Explain models and predictions through an adversarial approach  &
    \begin{minipage}[t]{\linewidth}
    \begin{itemize}[leftmargin=*]

      \item Methodology explaining incorrectly classified samples to help improve flaws in the model
    \end{itemize}
    \vspace{1mm}
    \end{minipage} &
    \begin{minipage}[t]{\linewidth}
    \begin{itemize}[leftmargin=*]

        \item Only tested on DoS attacks from NSL-KDD
    \end{itemize}
    \vspace{1mm}
    \end{minipage}
     \\
    \hline

    Feature-Oriented Design of Visual Analytics System for Interpretable Deep Learning Based Intrusion Detection \cite{wu2020feature} &

    A suite of visual tools used to improve explainability of CNNs  &
    \begin{minipage}[t]{\linewidth}
    \begin{itemize}[leftmargin=*]

      \item Analysis of Features and Requirements to improve visual analysis of XAI
      \item IDSBoard, a GUI for understanding Deep Learning Intrusion Detection
      \item Demonstrate the effectiveness of visual analytics
    \end{itemize}
    \vspace{1mm}
    \end{minipage} &
    \begin{minipage}[t]{\linewidth}
    \begin{itemize}[leftmargin=*]

        \item Only tested on a single dataset
        \item Scalability of visual analytics system
        \item Visual analytics system only designed for CNN
    \end{itemize}
    \vspace{1mm}
    \end{minipage}
     \\
    \hline


    Explanation framework for Intrusion Detection \cite{burkart2021explanation} &

    Explaining IDS explanations using a Counterfactual technique.  &
    \begin{minipage}[t]{\linewidth}
    \begin{itemize}[leftmargin=*]

      \item Explaining classifications based on feature importance.
      \item Advice on how to change a classification to its desired result.
      \item Outline the decision process so that the user can simulate it themselves.
    \end{itemize}
    \vspace{1mm}
    \end{minipage} &
    \begin{minipage}[t]{\linewidth}
    \begin{itemize}[leftmargin=*]

        \item Analysis of the counterfactual technique was only run on one type of ML algorithm.
    \end{itemize}
    \vspace{1mm}
    \end{minipage}
     \\
    \hline
    \rowcolor{cyan!20!}
    \multicolumn{4}{c}{Decomposition/Gradient based IDS} \\
    \hline

    Toward Explainable Deep Neural Network Based Anomaly Detection. \cite{amarasinghe2018toward} &

    Initial steps into XAI for DNN Intrusion Detection &
    \begin{minipage}[t]{\linewidth}
    \begin{itemize}[leftmargin=*]

      \item Framework for creating an explainable Deep Network XAI concepts to improve trust
    \end{itemize}
    \vspace{1mm}
    \end{minipage} &
    \begin{minipage}[t]{\linewidth}
    \begin{itemize}[leftmargin=*]

        \item Experiments are run using only DOS attacks from the NSL-KDD dataset
    \end{itemize}
    \vspace{1mm}
    \end{minipage}
     \\
    \hline
    
\end{tabularx}
}

\caption{An overview of the existing literature on black-box approaches to intrusion detection systems, with a focus on their scope, contribution, and limitations. (Continued)} 
\label{table:blackbox_table}

\end{table*}

\addtocounter{table}{-1}

\begin{table*}
\centering
{
    
    {\renewcommand{\arraystretch}{1.30}%
    \begin{tabularx}{\textwidth}{X|X|X|X}
    \hline

    Towards explaining anomalies: A deep Taylor decomposition of one-class models. \cite{Kauffmann2020TowardsEA} &

    Explaining anomalies found by a SVM using Deep Taylor Decomposition.  &
    \begin{minipage}[t]{\linewidth}
    \begin{itemize}[leftmargin=*]

      \item A method for `neuralizing' a one-class SVM to be explained by Deep Taylor Decomposition.
    \end{itemize}
    \vspace{1mm}
    \end{minipage} &
    \begin{minipage}[t]{\linewidth}
    \begin{itemize}[leftmargin=*]

        \item Experiments solely run using one-class SVM. No comparison to `real' neural networks.
    \end{itemize}
    \vspace{1mm}
    \end{minipage}
     \\
    \hline
  
   \rowcolor{cyan!20!}
   \multicolumn{4}{c}{Hybrid IDS} \\
   \hline
    Achieving explainability of intrusion detection system by hybrid oracle-explainer approach  \cite{szczepanski2020achieving} &
    Building a hybrid IDS based around `XAI Desiderata' that does not decrease performance or add vulnerability.  &
    \begin{minipage}[t]{\linewidth}
    \begin{itemize}[leftmargin=*]
      \item An explainer module modeled after the `XAI Desiderata.'
      \item A Hybrid-Oracle explainer Intrusion Detection System.
    \end{itemize}
    \vspace{1mm}
    \end{minipage} &
    \begin{minipage}[t]{\linewidth}
    \begin{itemize}[leftmargin=*]

        \item Two models need to be effectively trained.
        \item Would benefit from being tested on multiple datasets.
    \end{itemize}
    \vspace{1mm}
    \end{minipage}
     \\
    \hline
  
      
    Explainable deep few-shot anomaly detection with deviation networks \cite{pang2021explainable} &
    An anomaly detection system able of detecting anomalies learned from few anomalous training samples.  &
    \begin{minipage}[t]{\linewidth}
    \begin{itemize}[leftmargin=*]
      \item Prior-driven anomaly detection framework.
      \item DevNet, an anomaly detection framework based on Gausian prior, Z-Score-based deviation loss, and multiple instance learning.
      \item A theoretical and empirical analysis of Few-shot anomaly detection.
    \end{itemize}
    \vspace{1mm}
    \end{minipage} &
    \begin{minipage}[t]{\linewidth}
    \begin{itemize}[leftmargin=*]
        \item Experiments only run using image based datasets with relatively small sample sizes.
    \end{itemize}
    \vspace{1mm}
    \end{minipage}
     \\
    \hline 
    \end{tabularx}
}
\caption{An overview of the existing literature on black-box approaches to intrusion detection systems, with a focus on their scope, contribution, and limitations. (Continued)}}
\end{table*}

\subsubsection{Feature based approaches}

One popular scheme for explanations considers the influence features have on prediction. Such schemes are called feature explanations. Existing processes, such as feature engineering and feature selection, are already common in machine learning pipelines. Therefore, it is natural that features emerge as a method of explainability. Several candidate solutions that currently exploit this assumption are Partial Dependence Plot (PDP), Accumulated Local Effects (ALE), H-statistic, and SHAP.



An important generalizable SHAP-based framework is proposed by Wang et al. \cite{wang2020explainable}. Their framework uses both local and global explanations to  increase the explainability of the IDS model.   
The IDS model consists of a binary Neural Network (NN) classifier and a multi-class NN classifier. To generate explanations, both models and predictions are fed to the SHAP module. 
Local explanations are generated by choosing an attack and randomly selecting 100 of the occurrences. An average Shapely value is calculated, and the SHAP module outputs a confidence score for the prediction. 
The authors evaluate explainability by using a neptune attack, where a flooding of SYN packets is observed. The explanation results show that the top four features are related to DoS and SYN flood attacks. Using the global explanation produced by the SHAP module, researchers can make inferences about how the model might react during a related attack. 
However, the model's confidence seems to favor attacks that attempt many network connections (e.g. probe or DoS) over other attacks, such as privilege escalation attacks. 
The IDS system along with the SHAP explanations are relevant to assist subject matter experts in making security decisions.

In another effort, Islam et al. \cite{islam2019domain} built a domain knowledge infused explainable IDS framework. Their architecture is composed of two parts: a feature generalizer that uses the CIA principles and an evaluator that compares the black box models using different configurations. 

The feature generalizer first maps the top three ranked features to attack types, then maps attack types to the CIA principles. 
For example, DoS attacks are associated with availability; Heartbleed or PortScan attacks are associated with confidentiality. 
Using this mapping system, the authors add three new features: C, I, and A. These three features include the aggregate scores of their related features from a data sample. If a feature positively affects a prediction, then it adds to the score; otherwise, it subtracts from the score. 

The evaluator, on the other hand, runs four different feature configurations. The first configuration uses the full, preprocessed CICIDS dataset of 78 features. The second is a feature selected dataset of 50 attributes. The final two datasets are domain knowledge based: a 22 feature dataset of domain infused features and a three feature dataset consisting of C, I, and A scoring features.  

Tests are run on ANN, SVM, Random Forest (RF), Extra Trees (ET), Gradient Boosting (GB), and Naive Bayes algorithms (NB). The authors outline two types of tests: explainability and generalizability. The first two datasets are used to find a baseline to compare against the authors' novel, domain infused approach. Their findings from initial experimentation show that the RF using the full dataset outperforms all other algorithms with an F1-score of 99.68\%. The domain infused and CIA datasets are able to obtain an F1-score of 99.32\% and 93.84\% on RF and ET algorithms, respectfully. The small difference between the full dataset and the domain infused dataset show that the authors can now create a way to explain predictions without negatively impacting model performance. The authors create another CIA scoring formula that shows how much impact a CIA mapped feature had on the samples prediction. These C, I, and A scores can then be shown to an analyst to explain the prediction. To test their method against unknown attacks, the models are trained on all attacks in the dataset except one. The classifier is tested on a dataset that includes all of the attacks. 

The results show that the novel domain infused dataset performs similarly to the full dataset. In one case, the domain infused dataset is able to be used to find an attack that the full dataset configuration could not. The authors have demonstrated that creating an explainable algorithm and dataset can be useful for both accuracy and explanations. 

Sarhan et al. proposes another feature based technique \cite{Sarhan2021AnEM}. Two feature sets, NetFlow and CICFlowMeter, are evaluated across three datasets. When new IDS datasets are created, they are not necessarily created using the same tools. NetFlow and CICFlowMeter based IDS datasets collect different feature sets. The authors test these different feature sets using Random Forests and Deep Feed Forward algorithms. The results from this experiment show a minor improvement from the NetFlow feature set over the CICFlowMeter set. The most interesting result is the change in false positives between the two feature sets. NetFlow offers a much lower false positive rate than its counterpart in many of the tests. Additionally, NetFlow is slightly faster to make predictions than CICFlowMeter. The authors conclude that NetFlow offers slightly higher quality security features. Explainability is achieved in the form of SHAP. SHAP is used to determine which features are causing this difference in performance. The authors conclude that there are certain features across all datasets that contain more security focused data. However, the most important features vary across datasets. This is attributed to the fact that each dataset has different attacks. The authors work shows the importance of feature selection during dataset creation.

A novel method in \cite{Antwarg2019ExplainingAD} uses Auto-Encoders (AE) in combination with SHAP to explain anomalies. Anomalies are detected using the reconstruction score of the AE. Samples that return a higher reconstruction score are considered anomalous. An explainer module is created with the goal to link the input value of anomalies to their high reconstruction score. Features are split into two sets. The first set contains features that are causing the reconstruction score to be higher, while the second set does the opposite. The authors label these sets `contributing' and `offsetting', respectively. Contributing features will have a SHAP score that is negative, and the opposite is true for offsetting. Explanations are presented in the form of a color-coded table where darker values are more important than lighter values. This novel approach to explaining AE can be improved with more iterations of its visualization style and methodology.

\subsubsection{Perturbation based approaches} \label{subsubsection:perturbation_based} 
Perturbation based approaches make minor modifications to input data to observe changes in output predictions. Their explanations are based on the inclusion, removal, or modification of a feature in a dataset. These approaches are model agnostic (see Section \ref{subsection:model_agnostic}), therefore, they can be applied to any model.



A representative work by Wu et al. \cite{wu2020feature} showcases the advantages of this approach. 
The authors have created a CNN model along with a dashboard user interface (UI) to make the black box deep learning components more explainable. They gather feature requirements for their dashboard from literature. These include: (i) it is important to know the role that individual neurons play in predictions; (ii) multiple models should be tested, and the best parameters should be selected to achieve the best accuracy; (iii) visualization should assist in finding interesting results; (iv) there should be an explanation as to how the model made a decision; (v) we should be able to see the data representation in each layer of the model. 

The authors use the NSL-KDD dataset to test their CNN. NSL-KDD is encoded into a 12x12 grayscale image that serves as input. Their model is able to achieve an 80\% accuracy. The dashboard UI is able to showcase a variety of visualizations that assists in explainability. The UI includes: a detailed view of each cluster of neurons and the associated feature class, a t-SNE scatterplot of the activation values, a feature map of the convolutional kernel, a feature panel that explains how the model came to a prediction (utilizing LIME and a Saliency chart), a confusion matrix of predicted instances, and a graph for finding input data patterns. The authors demonstrate the advantages of using the dashboard UI by comparing CNNs with fewer layers than their proposed architecture. For example, the last layer in a smaller CNN shows that it is unable to detect one of the attack types (u2r) from the NSL-KDD dataset, while the proposed architecture can detect the attack. The dashboard UI is able to demonstrate that the smaller model may need more layers to be effective. 

Khan et al. \cite{khan2021new} propose an explainable autoencoder-based detection framework using convolutional and recurrent networks to discover cyber threats in IoT networks. The model is capable of detecting both known and zero-day attacks. It leverages a 2-step, sliding window technique that is used to transform a 1-dimensional (1D) sample into smaller contiguous 2-dimensional (2D) samples. This 2D sample is then fed through a CNN, comprised of a 1D convolutional layer and a 1D max-pooling layer which extracts spatial features. The data is then fed into the auto-encoder based LSTM that extracts temporal features. Finally, the DNN uses the extracted representation to make predictions. To make the model explainable, the authors use LIME \cite{ribeiro2016should} (see Section \ref{localxai}). The dataset used for experimentation was from a real-world gas pipeline system. It consists of system logs that include packet data used to communicate with the pipeline, along with features such as packet length, pressure setpoint, and PID gain. The authors obtain a 99.35\% accuracy using their proposed model. LIME shows that there are five features in the dataset that are primarily responsible for the different predictions.

In another impactful work \cite{marino2018adversarial}, the authors ague that rather than explaining every prediction, it is possible to create a model that explains misclassifications using a \textit{counterfactual technique}. The goal  is to explain adversarial attacks, which aim to confuse models into misclassifying input samples. Using this technique, the authors find weak points in their model and develop strategies to overcome these limitations. When an input sample is classified incorrectly, minimal changes are made to the sample until it is classified correctly. The difference between the original, incorrectly labeled sample and the new, correctly labeled sample are used to explain the occurrence of the misclassification. 

NSL-KDD dataset is used to create these models. A linear classifier and a multi-layer perceptron (MLP) are used during testing and the authors achieve an accuracy of 93\% and 95\%, respectfully. t-SNE is used to visualize the misclassified and corrected samples. The authors technique for minimizing the difference between samples is effective as the projections created by t-SNE are nearly identical. More insight can then be gathered from these projections as they show which features caused the misclassification along with the magnitude of the impact. This method appears to be a good way to communicate why a classification occurred and allows for a user to make the necessary inferences.


Burkart et al. \cite{burkart2021explanation} proposes a similar application of counterfactuals on an explainable IDS framework. Here the goal of the system is to answer the question: \textit{Why did X happen and not Y?} The authors aim to create explanations that are \textit{understandable} and \textit{actionable}. By understandable, they mean explaining an instance of classification, and by actionable they mean giving advice for changing the classification. The framework should also allow the users to simulate these changes themselves. The counterfactual technique is used to achieve these goals.


The technique takes a vector $x$ and locates a similar co-ordinate position $x\prime$ in the feature space, that causes a change in the predicted label. $x\prime$ should be a sample that is very similar to \textit{x}. The authors' method for their explainable technique has 5 phases. In Phase 1, their algorithm finds the first counterfactual point by using an optimization problem. Phase 2 extrapolates that point by finding other points near it that are also opposite of the original vector $x$. By adding more than one counterfactual point, the algorithm can help find a better general understanding of the feature area. In their approach, the authors use \textit{MagneticSampling} to achieve this goal. This set of points is used in Phase 3 to find the decision boundary. Phase 4 takes this approximated decision boundary and trains a \textit{surrogate explainer model} for samples on both sides of the decision boundary. Phase 5 is the culmination of all of the previous work resulting in explanations, which include a feature importance explanation, a relative difference explanation, and a surrogate visualization module. The surrogate visualization can be done in a variety of ways; however, the authors choose to use a white box decision tree. The fidelity of their explainer is tested against LIME. Additionally, their method tests 2 varieties of explainers: a decision tree explainer and a linear explainer similar to LIME. The authors method performs better than LIME when \textit{MagneticSampling} is used in Phase 2, but performs worse than LIME when random sampling is used. The tree performs better than the linear method and the authors believe that it is superior based on its performance and inbuilt explainability.

\subsubsection{Decomposition Based Approaches} \label{subsubsection:decomposition_based}
Decomposition based approaches decompose the output of a model to create a relevance score. Layer-wise Relevance Propagation (LRP) is a technique where the scoring mechanism propagates backwards from the output node, highlighting activated neurons that impact predictions. According to the authors in \cite{arras2017explaining}, these approaches can either decompose the output or decompose the gradient of the model.
  \begin{figure}[h!]
    \centerline{\includegraphics[scale=0.80]{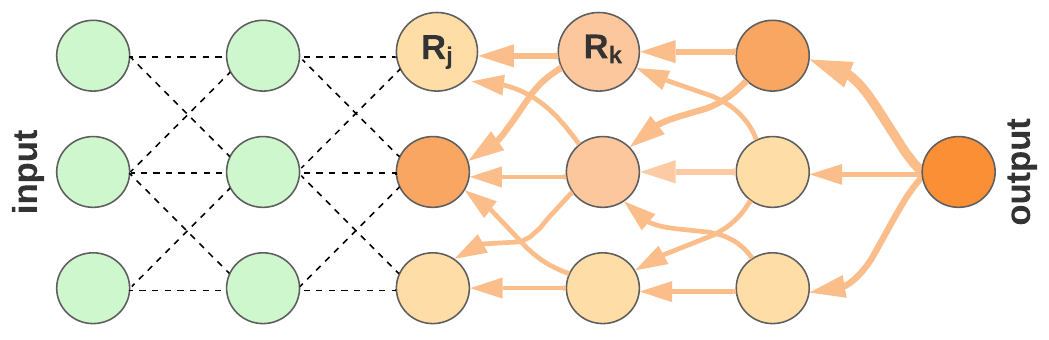}}
       \caption{A visual depiction of Layer-wise Relevance Propagation. Relevance scores (R$_j$, R$_k$) are calculated backwards from the output for each layer (\textit{j} and \textit{k} represent neurons). Scores from each previous layer are used to score the next set of neurons with the final outcome being the importance of each input \cite{montavon2019layer}.}
    \label{fig: lrp}
\end{figure}

An explainable DNN using LRP has been proposed in \cite{amarasinghe2018toward}. The goal of this system is to give a confidence score of a prediction, give a textual explanation of a prediction, and the reasons why the prediction was chosen. Online, a user can see that an anomaly has been found and why it is considered an anomaly, while offline an expert can evaluate the explanations. The authors argue that the explanation for detected anomalies is provided to reduce the `opaqueness' of DNN model and enhance `human trust’ in the algorithm. For their experiment, the authors create a partial implementation of their framework consisting of a Feed Forward DNN with explanations created by LRP. NSL-KDD is used for their experimentation. The tests are run using 4 different DNN configuration: two with three hidden layers and two with four hidden layers. Additionally, the dataset was separated into a `simple' dataset (a smaller number of features) and a `complete' dataset (all the features). The authors were able to achieve up to 97\% accuracy from each of the implementations. The model performed better with the complete dataset rather than with the simple dataset. The authors argue that the explainability of the simple dataset is worse than the complete dataset. This is because LRP chose a feature that would be difficult for a domain expert to verify. For example, binary features like `flag' are more difficult to explain than continuous features. The most important features for the complete dataset contained continuous values that could more easily be determined to be anomalous (src byte count and destination host count). Although the authors do not create a complete implementation with full textual explanations, their methodology could prove useful to improving the trust of regular users.

To address the issue of decomposability of DL models, the authors in \cite{Caforio2021LeveragingGT} propose an IDS system, based on CNNs, called GRACE (GRad-CAM-enhAnced Convolution neural nEtwork). They generate visual explanations for CNN decisions by utilizing the Gradient-weighted Class Activation Mapping (Grad-CAM) \cite{ selvaraju2017grad}.

The authors use three different datasets including KDD-CUP-99, NSL-KDDCUP99, and UNSW-NB15 to train their model. The textual dataset is transformed using image encoding which converts the training sample from the 1D feature vector
form $\mathbf{X}^{1D}$ with size 1 $\times M$ to the 2D image form $\mathbf{X}^{2D}$ with size m $\times m$ (with $M  \leq m_{2}$) and fed to the CNN model. The final convolution layer of 2D CNN is used to create heatmaps of class activations on input images, i.e., 2D grids of scores. Each pixel in the grid represents traffic characteristics, e.g., Destination port (X1) or idle max(X77). The scoring mechanism demonstrates how important each pixel (feature) is to a specific output class. This understanding of the most important features aids in the feature engineering process, resulting in a CNN model with higher accuracy.

To evaluate the performance of the model three evaluation metrics are used such as F1-Score (F1), Accuracy (A), and Computational complexity (T) (time spent to train the model). Of these, F1 and A metrics are used to compare against state-of-the-art such as CNN, GAN, LSTM, RNN, Triplet, DNN, MLP, and Autoencoder. Experimental results suggest GRACE generally outperforms its competitors. However, there are a few exceptions where the proposed model suffers slightly. For instance, when using the NSL-KDD dataset, the Triplet methods obtain 86.6\% and 87.0\% of A and F1, respectively, compared against 85.7\% and 86.8\% of the proposed model. The authors argue that this explanation approach can aid in the development of a more robust intrusion detection model.

Kauffmann et al. \cite{Kauffmann2020TheCH} propose another decomposition strategy aimed at verifying that a `Clever Hans' strategy has not been adopted by the ML model. LRP is leveraged as an explainer module to aid in discovering this phenomenon. Three separate models are trained: a kernel density estimator, an autoencoder, and a deep one-class model. Image based anomaly detection datasets MNIST-C and MVTec are used for this experiment. A `Clever Hans' score is adopted that is simply the difference between the detection accuracy and explanation accuracy. Detection accuracy is the ROC score, and explanation accuracy is the cosine similarity between the ground-truth and the pixel-wise explanation. It renders a score between 1 and -1 where 1 expresses a `Clever Hans' phenomenon. Results from their testing show that, based on their scoring system, all of the models show some form of `Clever Hans' logic. To address this problem, the authors propose a method of bagging anomaly detectors. This solution does not remove the phenomenon, but it does help to reduce it.

The previous authors also explore Deep Taylor Decomposition (DTD) for model explainability \cite{Kauffmann2020TowardsEA}. DTD is a technique that decomposes each neuron in each layer to determine feature relevance. A `neuralized', one-class SVM is proposed that can be explained using DTD. The `neuralized' form is a mapping of distance between the original sample and the SVM created support vectors as the first layer. The second layer is a soft min-pooling layer that calculates the `outlierness' of samples. Samples can then be explained using DTD by decomposing each of the neurons in the prediction. In their experiment, they use image based datasets for finding anomalies. DTD is used to highlight anomalous pixels in each image.

\subsubsection{Hybrid Approaches}
A hybrid black box predictor, white box explainer has been created by Szczepanski et al. \cite{szczepanski2020achieving}. Their framework is built with principles from the ``XAI Desiderata”: Fidelity, Understandability, Sufficiency, Low Construction Overhead, and Efficiency \cite{hansen2019interpretability}. The authors aim to contribute a system that is reliable, easy to understand, flexible, and meets all previous criteria without losing accuracy. With these goals in mind, a framework that uses local explanations is created. Their framework includes an ANN that predicts samples and a white box explainer that takes the output of the ANN and the original sample as input. The explainer is model agnostic and replaceable with any other explanation algorithm. The authors' explainer uses a clustering algorithm that uses a heuristic called Mean Distance to Average Vector. Clustering is done based on all of the attributes except the label. $n$ centroids are computed for all features, then a model is trained for each centroid cluster created. Another distance based algorithm is used to find a centroid cluster that is both close to the predicted sample and gives the same prediction as the ANN. The selected cluster is then used as a visualized explanation for a prediction. The authors note that it is possible that the explainer may not return a valid tree and that the model should be trained on a feature rich, diverse dataset. The authors experiment using the CICIDS2017 dataset. The ANN is able to achieve an accuracy of 98\%, and the explainer is able to achieve an accuracy of 99\% with 200 clusters. The authors have created a system where there are effectively two predictors that are used to confirm and explain the other's prediction. 

Pang et al. \cite{pang2021explainable} create a framework based on Few-Shot Anomaly Detection (FSAD). The authors claim that their framework is interpretable and explainable through a probability based scoring method and an image demonstrating anomalous areas found in samples. One of the problems faced in IDS/Anomaly Detection is that models are generally trained on unsupervised, normal data. This makes it difficult for models to discern from normal and anomalous data. The authors aim using FSAD to improve detection rates. However, FSAD has difficulties learning a generalized representation of anomalies from a few samples and it is challenging to learn a robust representation of data with respect to anomolous data. To resolve this, the framework needs to be able to learn about anomalous samples but not learn that all anomalies are the same as the training samples. The authors achieve this by using a prior driven anomaly score and end-to-end optimization of anomaly scores with deviation learning based on the prior probability. The architecture of DevNet is composed of an Anomaly Scoring Network and a Reference Score Generator that outputs into a Multiple-Instance-Learning-based (MIL) deviation loss Score Learner. The Anomaly Scoring Network is a function $\phi$ that creates a scalar anomaly score for pieces of an input. In this case, the pieces of an input are parts of an image. The Reference Score Generator creates a reference score $\mu_r$, which is a mean score of randomly selected non-anomalous samples. The reference score is derived from a prior $F$. The function $\phi(X)$, $\mu_r$, and the standard deviation of $\mu_r$ are provided as input into the MIL Deviation Loss Learner whereby the goal is to optimize anomaly scores so that anomalies deviate significantly from normal samples. 

The framework is tested on a variety of image datasets for identifying defects, planetary bodies, and medical anomalies. DevNet is tested against five other models and performs better on 7 out of 9 datasets. DevNet is able to achieve an AUC score between 80\% to 98\% amongst all of the datasets. As for explainability, the authors demonstrate that the algorithm can display the anomalous region on an image. DevNet generates both a black-white image of the location of the defect and an overlaid image showing where the defect lies on the original image.

\subsection{White Box X-IDS Models}


Models that can provide an explanation to expert users without utilizing additional models are referred to as \textit{interpretable} or \textit{white box models} \cite{loyola2019black}. A white box model’s internal logic and programming steps are completely transparent, resulting in an interpretable decision process \cite{pintelas2020grey}. However, when the model is to be explained to non-expert users, it may demand post-hoc explainability, such as visualizations \cite{arrieta2020explainable}. This interpretability, on the other hand, usually comes at a price in terms of performance \cite{islam2021explainable}.


A myriad of white box approaches are available for intrusion detection. Our survey will focus on the approaches most commonly used in the literature, as per our overview presented in Section \ref{overview}  and the taxonomy showcased in Figure \ref{tax_fig}. Table \ref{table:existing_research_white_box} summarizes state-of-the-art research, challenges, and
contributions with respect to white box approaches for intrusion detection systems. 

\begin{table*}
\centering
{
   {\renewcommand{\arraystretch}{1.30}%
    \begin{tabularx}{\textwidth}{X|X|X|X}
    \hline
    \rowcolor{lightgray!20!}
    \textbf{Paper Title} & 
    \textbf{Focus/Objective} & 
    \textbf{Contribution} & 
    \textbf{Limitation} \\
    \hline
    \rowcolor{cyan!20!}
    \multicolumn{4}{c}{Regression based IDS} \tabularnewline
    \hline
    Explainable Machine Learning for Intrusion Detection via Hardware Performance Counters \cite{kuruvila2022explainable} &
    
    To develop an explainable X-IDS technique based on the double RR technique and utilizing HPC as a feature. &
    
    \begin{minipage}[t]{\linewidth}
    \begin{itemize}[leftmargin=*]
    
        \item Proposes an explainable HPC-based Double Regression (HPCDR) framework for intrusion detection with human-interpretable results.
        \item HPCDR is evaluated against real-world malware to determine whether it provides transparent hardware-assisted malware detection and to detect microarchitectural attacks with an indication of the malicious origin.
    \end{itemize} 
    \vspace{1mm}
    \end{minipage} &
    \begin{minipage}[t]{\linewidth}
    \begin{itemize}[leftmargin=*]
    
        \item DL models were not chosen to evaluate the optimal ML model. 
        \item Only Four HPC features were chosen for experimentation. 
        \item Other microarchitectural attacks (e.g. Prime+Probe) and malware (e.g. Rootkits) are not considered in the study. 
    \end{itemize} 
    \vspace{1mm}
    \end{minipage} \\
  
    \hline 
    \rowcolor{cyan!20!}
     \multicolumn{4}{c}{Decision Tree and Rule based IDS} \tabularnewline
    \hline
    
   XAI to Enhance Trust Management in Intrusion Detection Systems Using Decision Tree Model \cite{mahbooba2021explainable} & 
   
   Focused on the interpretability in a widely used benchmark dataset KDD datasets. &
    \begin{minipage}[t]{\linewidth}
    \begin{itemize}[leftmargin=*]
        \item Addressed XAI concept to enhance trust management that human expert can understand. 
        \item Analyzed the importance of feature based on the entropy measure for intrusion detection.
        \item Interpreted the rules extracted from the DT approach for intrusion classification.
    \end{itemize} 
    \vspace{1mm}
    \end{minipage}
    &
    
    \begin{minipage}[t]{\linewidth}
    \begin{itemize}[leftmargin=*]
        \item Information gain in decision trees is biased in favor of those attributes with more levels. 
        \item This behavior might impact prediction performance.

    \end{itemize} 
    \vspace{1mm}
    \end{minipage} \\
   
    \hline

    A Hybrid Approach for an Interpretable and Explainable Intrusion Detection System \cite{dias2021hybrid} &
    To design interpretable and explainable hybrid intrusion detection system to achieve better and more long-lasting security.
     &
    \begin{minipage}[t]{\linewidth}
    \begin{itemize}[leftmargin=*]
        \item Providing an IDS that stands out for its ML support on populating the knowledge base.
        \item Focus on interpretability and explainability, since it justifies the suggested rules, and the diagnosis performed to each asset.
    \end{itemize} 
    \vspace{1mm}
    \end{minipage}
    &
    \begin{minipage}[t]{\linewidth}
    \begin{itemize}[leftmargin=*]
        \item DT only considered as ML model for system design. 
        \item Knowledge base is small. 
    \end{itemize} 
    \vspace{1mm}
    \end{minipage} \\
    
    \hline

    \rowcolor{cyan!20!}
       \multicolumn{4}{c}{Statistical and Probabilistic Models} \tabularnewline
    \hline
        A New Method for Flow-Based Network Intrusion Detection Using the Inverse Potts Model \cite{pontes2021new} &
    To develop a new method for flow-based network intrusion detection using inverse statistical method.   & 
    \begin{minipage}[t]{\linewidth}
    \begin{itemize}[leftmargin=*]
        \item Implementation of a naturally interpretable flow classifier based on the inverse Potts model to be employed in NIDS.
        \item Performance comparison with other ML based models using three datasets. 
    \end{itemize} 
    \vspace{1mm}
    \end{minipage} &
    \begin{minipage}[t]{\linewidth}
    \begin{itemize}[leftmargin=*]
        \item Only binary classification is considered in the approach.
        \item Applicability of the proposed methods in real world data. 
    \end{itemize} 
    \vspace{1mm}
    \end{minipage} \\
    \hline

    \rowcolor{cyan!20!}
     \multicolumn{4}{c}{Clustering based IDS} \tabularnewline
    \hline
    Explainable unsupervised machine learning for cyber-physical systems \cite{wickramasinghe2021explainable}
    & Propose a novel Explainable Unsupervised Machine Learning (XUnML) approach using the Self Organizing Map (SOM) algorithm. 
    & 
    \begin{minipage}[t]{\linewidth}
    \begin{itemize}[leftmargin=*]
        \item Brief overview of Supervised Machine Learning (SML), Unsupervised Machine Learning (UnML), and XAI. 
        \item Exploring initial desiderata towards Explainable UnML (XUnML), defining XUnML terminology based on the terminology used for XAI, and exploring the necessity
        of XUnML for CPSs.

    \end{itemize} 
    \vspace{1mm}
    \end{minipage} &
    \begin{minipage}[t]{\linewidth}
    \begin{itemize}[leftmargin=*]
        \item Only clustering method is used. 
    \end{itemize} 
    \vspace{1mm}
    \end{minipage} \\
    \hline

   
    ANNaBell Island: A 3D Color Hexagonal SOM for Visual Intrusion Detection \cite{langin2011annabell}
    & Provide explanation to the outputs of SOM models using color scheme and island landscape analogy for different network traffics.
    & 
    \begin{minipage}[t]{\linewidth}
    \begin{itemize}[leftmargin=*]
        \item Benign and malicious traffic is separated by color coding and zoning in the island.  
        \item Color and zone categorization of network traffic provides the explanation of the output. 

    \end{itemize} 
    \vspace{1mm}
    \end{minipage} 
    & 
    \begin{minipage}[t]{\linewidth}
    \begin{itemize}[leftmargin=*]
        \item It is not clear if the temporal map maintain same basic landscape or change over time. 
        \item The proposed map seems to be specific to the tested network only. 
    \end{itemize} 
    \vspace{1mm}
    \end{minipage} \\
    \hline 
    \end{tabularx}
}
\caption{A summary of the existing literature on white-box approaches to intrusion detection systems, with an emphasis on their scope, contribution, and limitations.}
\label{table:existing_research_white_box}
}
\end{table*}

\subsubsection{Regression} Linear Regression (LR) is a supervised ML technique that establishes a relationship between a dependent variable and independent variables by computing a \textit{best fit} line. The linearity of the learned relationship puts LR under the umbrella of interpretable models. Intrinsically interpretable models such as LR and Logistic Regression (LoR), meet the characteristics of transparent models (algorithmic transparency, decomposability, and simulatability) \cite{lipton2018mythos}. In LR, when the number of features is small, the weight or coefficient of the equation can be used for explaining predictions. 
The learned relationships are linear and can be expressed for a single instance \textit{i} as given in Equation \ref{eq:regression_equation}. 

\begin{equation}\label{eq:regression_equation}
\hat{y}=\beta_{0}+\beta_{1} X_{1}+\cdots+\beta_{n} X_{n}+\epsilon
\end{equation}
where, 
\begin{quote}
$\hat{y}$ is the output or target (dependent) variable. X is the input (independent) variable. $\beta_{0}$, $\beta_{1}$ are coefficients, and $\epsilon$ is  error term. 
\end{quote}

Various regression-based  IDS models exist in the literature. Authors in \cite{subba2015intrusion} deployed anomaly-based intrusion detection systems using two different statistical methods: Linear Discriminant Analysis (LDA) and LoR. 
While LR models are desirable for intrusion detection purposes, their performance is susceptible to outliers \cite{wang2014new}. To mitigate the impact of outliers, the authors in \cite{wang2017robust} proposed a robust regression method for anomaly detection. The proposed method uses heteroscedasticity and a huber loss function instead of homoscedasticity and sum of squared errors. 

While the existing approaches render promising outcomes, none of them were designed with \textit{explainability} in mind. To overcome the issue of \textit{explainability} in the area of \textit{hardware performance counter} (HPC) – based intrusion detection, the authors \cite{kuruvila2022explainable} propose an explainable HPC-based Double Regression (HPCDR) ML framework. The study examines two distinct types of attacks: microarchitectural and malware. For the first type of attack, tests are conducted on five distinct datasets: Rowhammer, Flush+Flush, Spectre, Meltdown, and ZombieLoad. For the second attack, two distinct datasets are considered: Bashlite and PNScan. To minimize computational overhead, the proposed study employs Ridge Regression (RR) rather than Shapely values to generate interpretable results. First, the three ML models (RF, DT, and NN) are chosen to evaluate the classification accuracy. Second, the output from these models is perturbed and passed to the first RR model where HPCs are employed as features and weight coefficients are received. These furnished coefficients are run on the second RR model, which identifies the most malicious sample. The authors argue that by utilizing double regression techniques, their proposed method provides transparency, which enables users to locate malicious instructions within the program. 

\subsubsection{Decision Tree and Rule Based} 
\begin{figure}[h!]
    \centerline{\includegraphics[scale =.55]{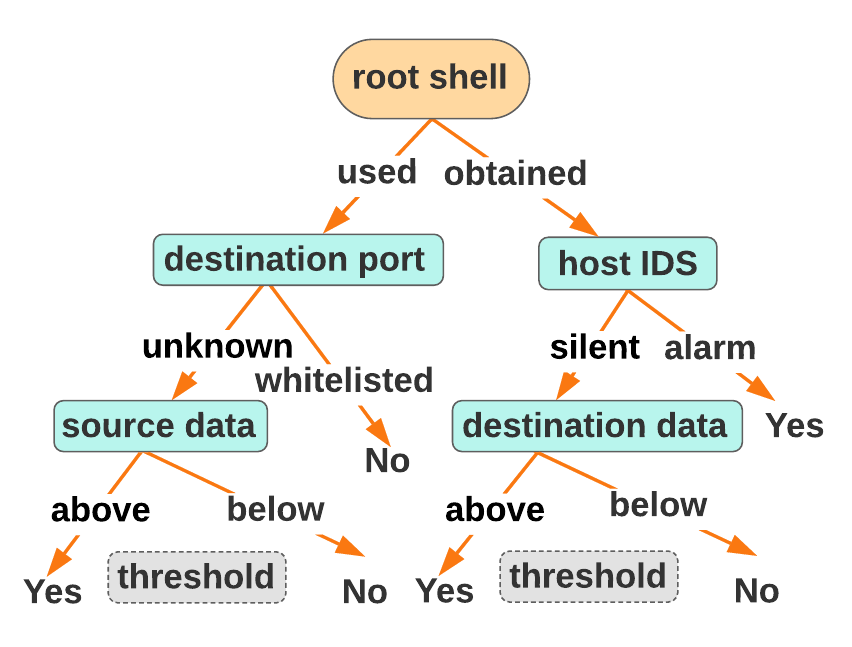}}
    \caption{Simple decision tree for the detection of a user-to-root (U2R) attack on a computer system, attack classified as Yes or No  \cite{staudemeyer2012importance}.}
    \label{fig: decision_tree}
\end{figure}

A Decision Tree (DT) is a tree structure with decision support system elements based on graph theory.  In contrast to LR and LoR methods, it works even when the relationship between input and output is nonlinear. There are two approaches for constructing decision trees: top-down and bottom-up. The top-down approach, which is based on the divide and conquer strategy, is the most frequently used in the literature \cite{quinlan2014c4}. The decision model built with a top-down approach begins with the root node and splits the root nodes into two disjoint subsets: left child and right child. This process is repeated recursively over child nodes until a stop condition is met \cite{maimon2014data}. In their simplest form, DT possesses three properties that make them interpretable \cite{lipton2018mythos}: simulatability, decomposability, and algorithmic transparency \cite{arrieta2020explainable}. Figure \ref{fig: decision_tree} illustrates a simple DT that detects an attack on a remotely accessible computer system.

\begin{figure}[h!]
    \centerline{\includegraphics[scale=0.55]{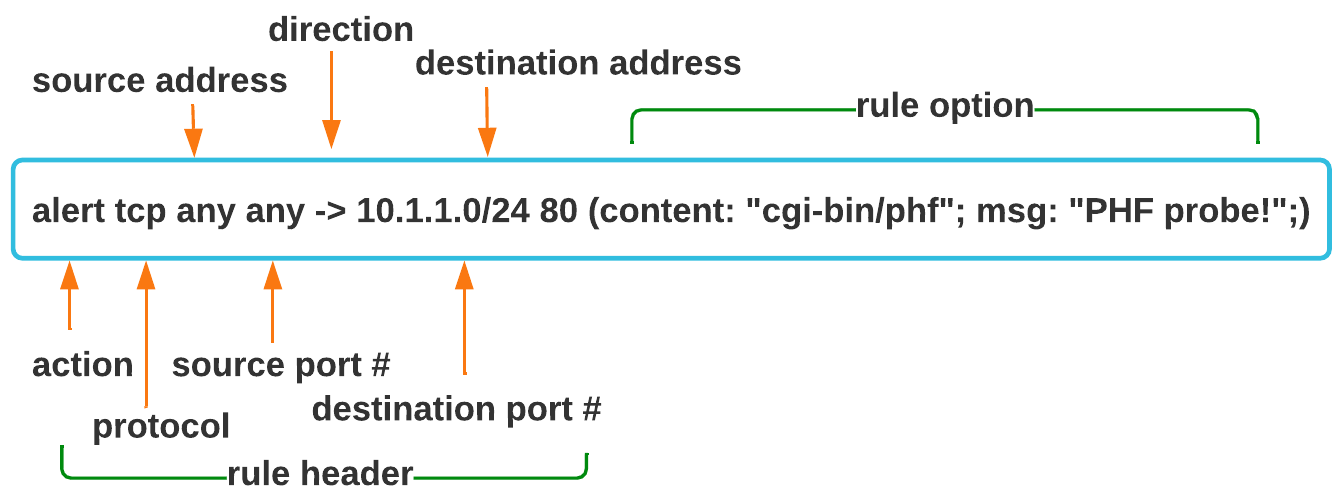}}
       \caption{A simple rule that detects attempts to access the PHF service via a local web server. Any time such a packet is detected on the network, an alert is sent and the entire packet is logged using the run-time logging mechanism  \cite{roesch1999snort}.}
    \label{fig: rule_based}
\end{figure}

A simple rule is typically represented as a logical implication of IF-THEN statements by combining relational statements to form their knowledge \cite{loyola2020explainable}. These rules can be extracted from DT. Rule-based models are considered transparent because they generate rules to explain their predictions. Figure \ref{fig: rule_based} depicts a simple snort rule.

Mahbooba et al. \cite{mahbooba2021explainable} approach the task of developing an interpretable model to identify malicious nodes for IDS using a DT on the KDD dataset. They chose the Iterative Dichotomiser 3 (ID3) algorithm to ensure interpretability because it mimics a human-based decision strategy. The authors demonstrate that the algorithm can rank the relevance of features, provide explainable rules, and reach a level of accuracy comparable to state-of-the-art. Another explainable decision tree model is proposed in \cite{loyola2020explainable} and \cite{frost2020exkmc}, with the latter being an extension of work in \cite{dasgupta2020explainable}.

Sinclair et al. \cite{sinclair1999application} extract rules using a DT and a Genetic Algorithm (GA) for improving the performance of the IDS model. The authors in \cite{ojugo2012genetic} and \cite{chadha2015hybrid} focus on optimizing the IDS model by extracting rules using a GA. To add transparency to the decision process, Dias et al. \cite{dias2021hybrid} proposed an interpretable and explainable hybrid intrusion detection system. The proposed system integrates expert-written rules and dynamic knowledge generated by a DT algorithm. The authors suggest that the model can achieve explainability through the justifications of each diagnosis. Justification of certain predictions is provided in a tree-like format in the form of a suggested rule that provides a more intuitive and straightforward understanding of the diagnosis.

\textit{Snort} is the world's most widely used open-source rule-based intrusion prevention system (IPS) \cite{roesch1999snort}. It employs a set of rules that help define malicious network activity. These rules are then used to identify packets and generate alerts for users \cite{roesch1999snort}, \cite{caswell2007snort}.

     \subsubsection{Statistical and Probabilistic Methods}
     \label{Statistical and Probabilistic Models}

    In statistics, the mean, standard deviation, and any other type of correlation are referred  to as moments \cite{qayyum2005taxonomy}, \cite{jyothsna2011review}. 
    Statistical and probabilistic methods use this information to determine whether the given event is anomalous or not. The moment is predicted anomalous if they are either above or below a predefined interval \cite{hodo2017shallow}.
 This approach is further divided into the univariate model, multivariate model, time series model in \cite{khraisat2019survey}, parametric and non-parametric model in \cite{bhuyan2013network}, \cite{tran2017network}, operational model, Markov model and statistical moments in \cite{qayyum2005taxonomy}, \cite{gyanchandani2012taxonomy}.

    Various IDS based on statistical and probabilistic models have been proposed. IDS based on the mean and standard deviation is explained in \cite{ashfaq2010information}, while a study relating to multivariate modeling is proposed in \cite{sha2013multi}. The authors in \cite{ye2000markov} proposed an IDS based on the Markov process.

    A different approach to intrinsically explainable statistical methods for network intrusion detection is proposed by Pontes et al. \cite{pontes2021new}. The authors in this study introduce a novel Energy-based Flow Classifier (EFC) that utilizes inverse Potts models to infer anomaly scores based on labeled benign examples. This method is capable of accurately performing binary flow classification on DDoS attacks. They perform experiments on three different datasets: CIDDS-001, CIC-IDS2017, and CICDDoS19. Results indicate that the proposed model is more adaptable to different data distributions than classical ML-based classifiers. Additionally, they argue that their model is naturally interpretable and that individual parameter values can be analyzed in detail.  
    

\subsubsection{Clustering}
Clustering is the most widely used strategy for unsupervised ML. It classifies samples according to a similarity criterion. 
Clustering algorithms that can be explained have several advantages. The primary benefit of explainable clustering is that it summarizes the input behavior patterns within clusters, enabling users to comprehend the clusters' underlying commonalities \cite{wickramasinghe2021explainable}. As stated in Section \ref{overview_whitebox_approach} there are various clustering algorithms available. However, in the context of X-IDS, we will only focus on Self-Organizing Maps (SOMs).

\begin{figure}[h!]
    \centerline{\includegraphics[scale =0.55 ]{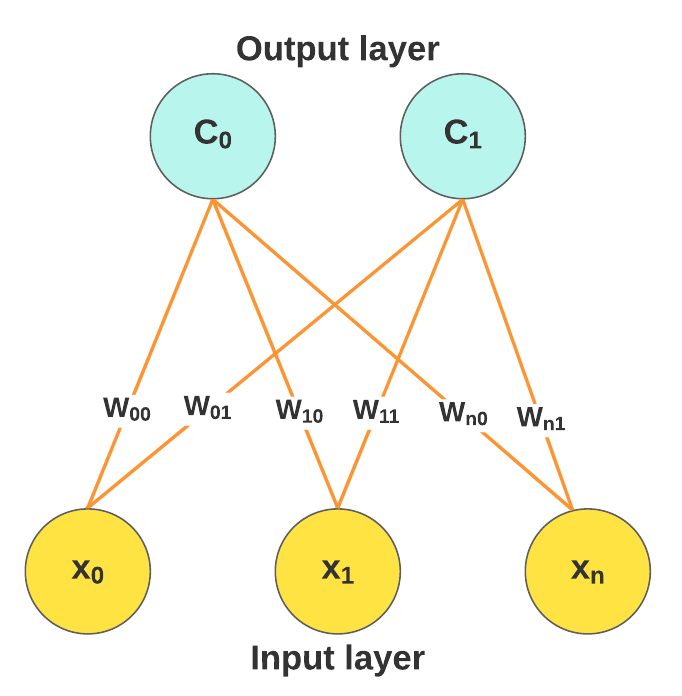}}
    \caption{Clustering based on Self Organizing Maps. The input layer consists of $x_{0}, x_{1}...x_{n}$ nodes. These nodes have specific weights illustrated by $w_{00}, w_{01},...w_{n1}$. These weight determines cluster classification such as $C_{0}$ and $C_{1}$.}
    \label{fig: SOM}
    \end{figure}

SOMs are an unsupervised clustering technique within the artificial neural networks umbrella. It has two layers: an input layer that accepts high dimensional space and an output layer that generates a non-linear mapping of high-dimensional space into reduced dimensions.  It is trained to produce a low dimensional representation of a large training dataset while preserving important topological and metric relationships of the input data \cite{kohonen1996engineering}. A graphical illustration of a simple SOM model is depicted in Figure \ref{fig: SOM}.

An anomaly detection system using SOM techniques based on offline audit trail data is proposed in \cite{hoglund2000computer}. The major shortcoming of the proposed system is it does not allow for real-time detection. On the other hand, the authors in \cite{lichodzijewski2002host} propose Hierarchical SOMs (HSOM) for host-based intrusion detection on computer networks that are capable of operating on real-time data without requiring extensive offline training or expert knowledge. Another model based on HSOM is proposed in \cite{kayacik2007hierarchical}. 
The authors in \cite{wickramasinghe2021explainable} developed a novel model-specific explainable technique for the SOM algorithm that generates both local and global explanations for Cyber-Physical Systems (CPS) security. They used the SOMs training approach (winner-take-all algorithm) together with visual data mining capabilities (Histograms, t-SNE, Heat Maps, and U-Matrix) of SOMs to make the algorithm explainable. 
   
A 3D color hexagonal SOM for visual intrusion detection called ANNaBell Island is proposed in \cite{langin2011annabell} which is an extension of 1D ANNaBell reported in \cite{ langin2008model, langin2009self}. To make the SOM process and its output explainable to the users, the authors designed a hexagonal SOM in a meta-hexagonal layout, referred to as an island, that graphically displayed features of network traffic. The output of the SOM model was used to create a color-separated 3D landscaped island that represents various types of network traffic, distinguishing between malicious and normal behavior.

After surveying the current black box and white box approaches to X-IDS, we propose in the next section, a generic explainable architecture with a user-centric approach for designing X-IDS that can accommodate a wide variety of scenarios and applications without adhering to a specification or technological solution.

\begin{figure*}[ht]
    \centering
    \includegraphics[scale=0.68]{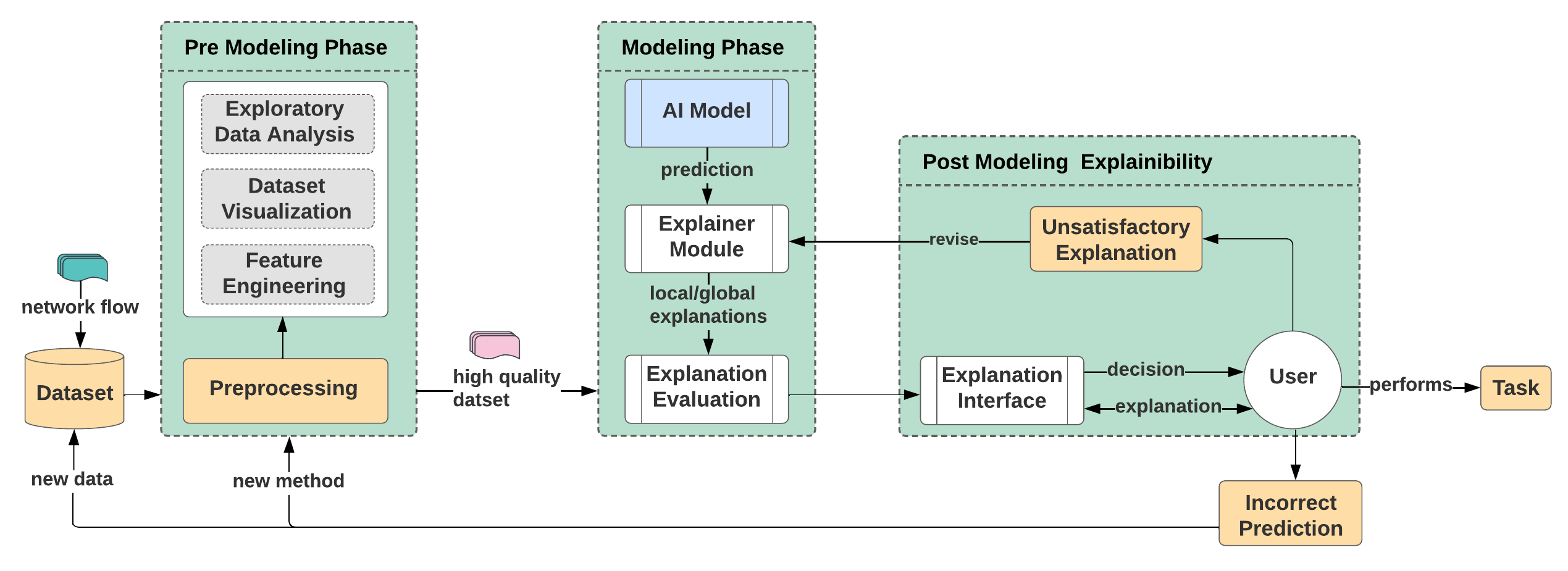}
    \caption{Recommended architecture for the design of an X-IDS based on DARPA \cite{gunning2019darpa}. The layered architecture is divided into three phases: pre-modeling, modeling, and post-modeling explainability. Each phase contributes to the development of an explanation for various stakeholders, thereby assisting in decision-making.}
        \label{fig:recommended_arch}
\end{figure*} 

\section{Designing an Explainable IDS (X-IDS)}
\label{design_xai_ids}


The purpose of an IDS is to continuously monitor a network for malicious activity or security violations known as incidents of intrusion. If found, intrusions are reported to the cybersecurity professional responsible for monitoring such systems. A significant problem with AI based IDS is their high false positive and false negative rates. Recently, many IDS based on ML/DL techniques have been proposed to address this issue, such as DNN \cite{amarasinghe2018improving, kim2017method}, RNN \cite{solch2015detecting, yin2017deep}, and CNN \cite{azizjon20201d, vinayakumar2017applying}. These techniques yield unprecedented detection accuracy. 
However, the effective use of these approaches require using high-quality data, as well as a considerable amount of computing resources \cite{holzinger2018machine}. Additionally, this modeling approach has typically suffered from model bias, a lack of decision process transparency, and a lack of user trust.

The IDS systems based on ML/DL techniques are designed to generate event logs in the form of `benign’ or `malicious’ classification reports, that can be further analyzed by CSoC analysts. However, they do not showcase the connection between the inputs and output (i.e., they fail to indicate the reasoning behind the decision). To be more precise,  a cybersecurity specialist serves as a user who reviews IDS results, but is not a component of the intrusion detection process \cite{liu2021faixid}. In turn, this creates a larger problem for CSoC experts, as they are unable to optimize their decisions based on the model's decision process.

To address this semantic gap, one promising technique is to design X-IDS with a human-in-the-loop approach. Typically, methods that are retraceable, explainable, and supported by visualizations amplify cybersecurity analysts’ understanding in managing cybersecurity incidents in both proactive and reactive manners.

In the following sub-sections, we explain the recommended architecture, as depicted in Figure \ref{fig:recommended_arch}, that could be used as guidance to design an Explainable Intrusion Detection Systems (X-IDS). The X-IDS architecture proposed in this paper is based on the DARPA recommended architecture for the design of XAI systems \cite{gunning2019darpa}. The layered architecture consists of three phases: \textit{pre-modeling phase}, \textit{modeling phase}, and \textit{post-modeling explainability phase}. In each phase, different modules work in tandem to provide CSoC analysts with more accurate and {explainable output}. We believe that this architecture is sufficiently generic to accommodate a variety of scenarios and applications without adhering to a particular specification or technological solution.

\subsection{Pre-modeling phase}

The first phase is a pre-modeling phase. The input of this module is raw network flow (dataset) and the output is a high-quality dataset. In the following subsections, we will first describe different benchmark datasets available for Intrusion Detection. We then present common data preprocessing techniques used in the literature.

\subsubsection{Datasets}
While access to representative, labelled datasets for cybersecurity related AI tasks remains a challenge, a variety of publicly accessible datasets can be used to train and benchmark X-IDS. These are unprocessed network flows extracted from packet captures. To address privacy concerns, many of these datasets are generated in an emulated environment. NSL-KDD \cite{tavallaee2009detailed}, based on the KDD-CUP-99 \cite{KDDCUP99}, is a dataset frequently present in the literature. 
Although old, its use allows comparisons with previous works. NSL-KDD is relatively small compared to other datasets in the field. A more modern dataset is CICIDS2017 \cite{sharafaldin2018toward}, which contains more up-to-date attacks and network flows. In addition, it includes 3 million samples, which allows scalability testing. Another noteworthy dataset is UGR \cite{macia2018ugr}, a multi-terabyte dataset collected over the course of 5 months. This dataset is built to test IDS for long-term trends. The authors state that their dataset captures potential trends in daytime, nighttime, weekday, and weekend traffic. 

These publicly available datasets, though good for benchmarking, are not suitable for deployable systems. We recommend, CSoC users deploying X-IDS systems evaluate these systems on organizational representative datasets.

\subsubsection{Exploratory Data Analysis (EDA) and Data Visualization:}

Data preprocessing is essential for increasing the likelihood of ML models producing accurate predictions. Using Exploratory Data Analysis (EDA), one can gain a general understanding of a dataset's key features and characteristics. To comprehend the features, visualization techniques such as heat maps, network diagrams, bar charts, and correlation matrices may be used. Once a comprehension of feature space has been attained, the data is forwarded to the feature engineering model for further processing.

\subsubsection{Feature Engineering}
The general trend in preprocessing IDS datasets is to normalize the numerical features and to One-Hot Encode (OHE) the categorical features. After the datasets are encoded, their feature space can be quite large which makes them computationally expensive. Two approaches to reducing dimensions are widely discussed in the literature: Feature Selection and Feature Extraction.

Feature selection techniques are used to reduce the feature space by selecting a subset of features without transforming them. There are three types of feature selection techniques popular in the IDS domain: filters, wrappers, and the embedded/hybrid method \cite{khalid2014survey}. Apart from these, libraries such as Scikit-Learn \cite{scikit-learn} have also been used in published works for feature selection. 

Another technique used in feature engineering is feature extraction (also known as dimensionality reduction).  Feature extraction reduces the size of the feature space by transforming the original features while retaining most of their defining attributes. The most commonly used feature extraction technique in the literature is the Principal Component Analysis (PCA) \cite{pearson1901liii}. PCA is an unsupervised method that does not require class knowledge to identify features. It also facilitates the identification of correlations and relationships between the features of a dataset.

\subsection{Modeling Phase}

The second phase is the modeling phase. The input of this phase is the high-quality dataset generated in the pre-processing phase and the output is the explanations generated by the explainer module. First, the high-quality dataset is fed into the ML/DL model of choice. Second, the predictions generated by the model in use are passed through an explainer module. Third, these explanations are evaluated by an evaluation module. This process enables users to understand the reason behind certain predictions, which in turn, helps the CSoC analysts in their decision-making process. 

\subsubsection{AI Model}
In Section \ref{xaiids}, we discussed two different approaches which are currently being employed by different authors to create X-IDS: the black box and the white box. AI modules in these approaches generate predictions. However, there is a trade-off between the accuracy and the interpretability with these approaches. The white box approaches are popular for their interpretability, while the black box approaches are known for their prediction accuracy. 
In context of IDS, high prediction accuracy is required to prevent attacks. 
Moreover, black box models can capture significant non-linearity and complex interactions between data that white box models are not able to capture. For example, Recurrent Neural Networks (RNN) can capture temporal dependencies between samples. On the other hand, models like Support Vector Machine (SVM) and Deep Neural Network (DNN) can create their own representation of data. which might be helpful to discover unknown attacks. For this reason, we believe that future X-IDS should be built using black box models.

In our literature review, we found that authors use a variety of black box algorithms in their work, such as SVM, CNN, RF, and MLP, which prove to be quite effective. Another popular algorithm of choice in the intrusion detection domain is a variant of the RNN, referred to as Long Short-Term Memory (LSTM). Recently, Generative Adversarial Networks (GAN) have also become relatively popular. Consequently, there are a multitude of black box algorithms from which to select. Explainer modules then approximate the prediction generated by AI module employing a white box or black box algorithms.


\subsubsection{Explainer Module and Evaluation} \label{exp_module_evaluation}
The prediction generated by the model of choice in the AI module is then fed to the explainer module. The common explainers used from previous works include LIME, SHAP, and LRP. These out-of-the-box modules allow for quick testing on different algorithms and datasets. However, there are some problems with solely using these approaches in future X-IDS works. To begin, methods such as SHAP do not run in real-time. Therefore, it may be time-consuming to attempt to use SHAP on a simple Multi-Layer Perceptron classifier with a large feature space dataset. In X-IDS, both predictions and explanations must be made as quickly as possible. Secondly, these approaches are not always designed with X-IDS stakeholders in mind. 

At present, there are no set standard metrics to evaluate explanations. Several authors have attempted to evaluate explanations in various ways. In Section \ref{exp_metric} we described different ways to evaluate the explanations. Metrics such as application grounded evaluation, human-grounded evaluations, and function-grounded evaluation proposed by Doshi et al. \cite{doshi2017towards} can be used as a baseline to evaluate the explanation generated by X-IDS. A noteworthy method to evaluate the effectiveness of explanations is proposed by authors in \cite{gunning2019darpa}. Figure \ref{fig: explainability_metrics}  illustrates their approach. 


\subsection{Post Modeling Explainability Phase}
The third and final phase is the post-modeling explainability phase. This phase has two major components: the explanation interface and users. The recommendation, decision, or action generated by the AI module, explained by an explainer module, and evaluated by an explanation evaluation module is rendered in a graphical user interface (explanation interface). The users, on the other hand, use this interface to make an informed decision. 

\subsubsection{Explanation Interface}
The custom visual dashboards are created to help the user to understand the X-IDS. An excellent approach to building such an explanation interface is found in the work by \cite{wu2020feature} and \cite{islam2019domain}. The engineers who design X-IDS can use this approach as guidance to create their explanation interface. Furthermore, this paper also recommends that future X-IDS developers make custom explainers built for specific stakeholders to help improve explainability.


\subsubsection{Users}
For this paper, the stakeholders will consist of developers, defense practitioners, and investors. Section \ref{recommendation_stakeholder} discusses the need for defining the identity of the stakeholders of an X-IDS system. The developers are tasked with creating, modifying, and maintaining the X-IDS. The defense practitioners guard the assets of the investors. Lastly, the investors make budgeting decisions for the benefit of the X-IDS system and other assets. These three audiences have distinct tasks and explainability requirements that must be addressed differently by the X-IDS. An explanation interface designed from the user's perspective can bridge this gap.

If an explanation is unclear or unhelpful, the stakeholders will need a way to voice that opinion. In such a situation, the developers can revise the explainer module. For the same reasons, incorrect predictions and explanations need to be corrected and updated. The developers or defense practitioners will then need to introduce new data to the model.  Moreover, a different method of data preprocessing may be required to augment the efficacy of the model.

To make the recommended X-IDS architecture as shown in Figure \ref{fig:recommended_arch} a reality, researchers need to study different aspects of the three phases. To this end, in the next section we discuss various challenges inherent in designing the proposed X-IDS architecture and make research recommendations aimed at effectively mitigating these challenges for future researchers interested in developing X-IDS.

\section{Research Challenges \& Recommendations}
\label{recommendation}

The sub-domain of explainable AI based Intrusion Detection Systems is still in its infancy. 
Researchers working on X-IDS must be made aware of the issues that hinder its development. The issues that we described in Section \ref{xai} such as finding the right notion of explainability, generating explanations from a stakeholder’s perspective, and lack of formal standard metrics to evaluate explanations are prevalent in the X-IDS domain as well. Existing X-IDS research is primarily focused on the goal of making algorithms explainable. Explanations are not being designed around stakeholders, and researchers need to quantify useful evaluation metrics. Apart from these challenges, issues pertaining to IDS may also pose a problem for X-IDS. 
There are many promising avenues of exploration, in this section we detail some existing research challenges and give our recommendations.

\subsection{Defining Explainability for Intrusion Detection}


The first problem faced by researchers designing X-IDS is the lack of consensus on the definition of explainability in IDS. The research community needs to agree on a common definition of explainability for IDS. To find common ground, we can leverage the foundational XAI definition proposed by DARPA \cite{darpa2016broad}. However, an X-IDS definition needs more security domain-specific elements. The inclusion of the CIA principles may be a good start for cementing a definition that combines aspects of cybersecurity and XAI. 

Questions relevant to the X-IDS that researchers need to answer include: ``What is explainability when used for intrusion detection?”, ``How do we effectively create explanations for IDS?”, and ``Who are we creating explanations for?”. Other questions such as  ``How can Confidentiality, Integrity, and Availability benefit from explanations?” and ``How do we categorize X-IDS algorithms?” should be reassessed by X-IDS researchers as well. Current work is extremely narrow in its scope and limits its objective to  explaining each sample in an IDS dataset. These works also do not consider the type of audience when building X-IDS. 


\subsection{Defining Tasks and Stakeholders} \label{recommendation_stakeholder}
\begin{figure}[htbp]
    \centering
    \includegraphics[scale=0.6]{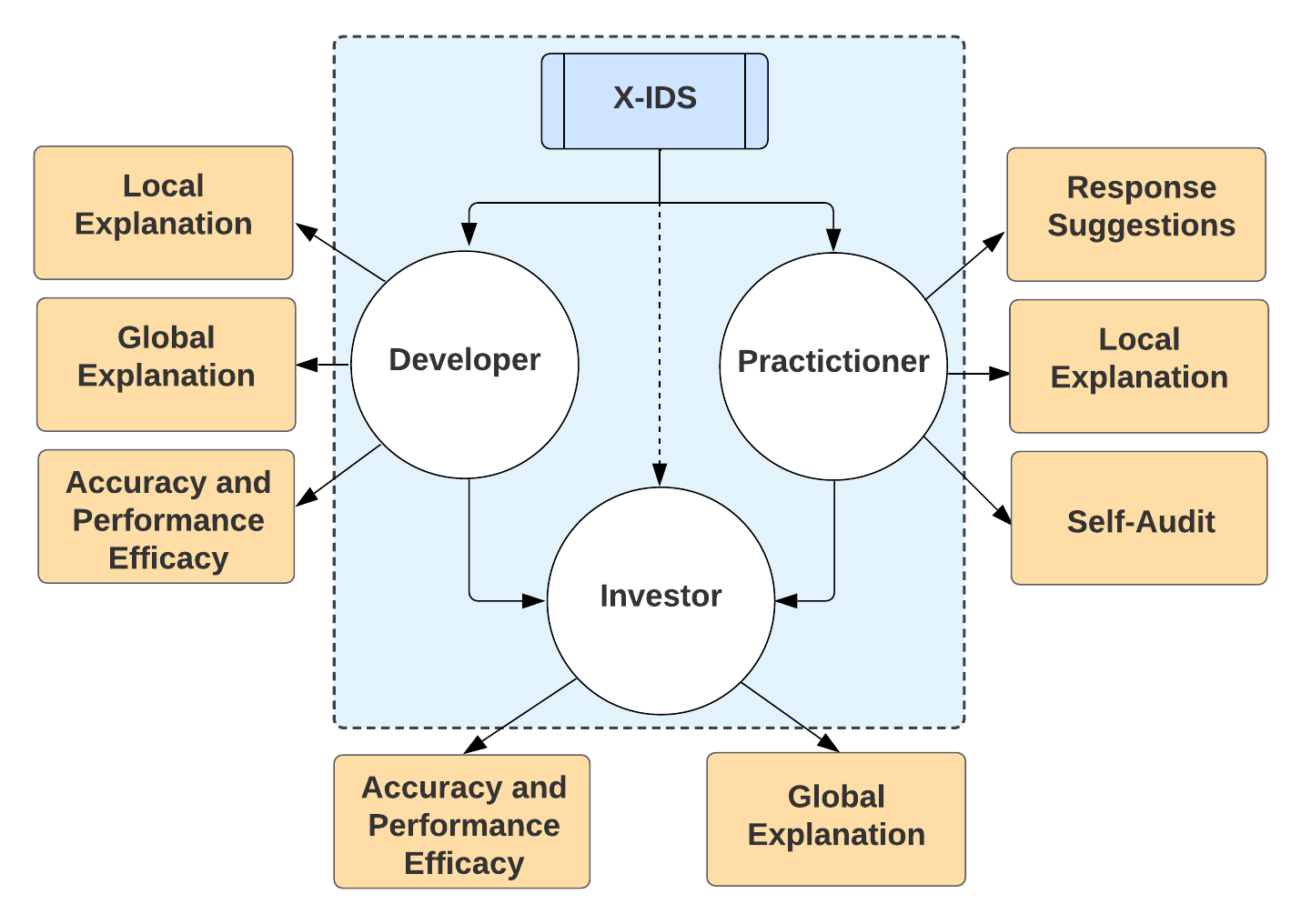}
    \caption{A simple taxonomy illustrating the importance of tailoring explanations to specific stakeholders based on their roles in CSoCs.}
        \label{fig: user_tax}
\end{figure}

The second challenge is to define the task and the stakeholders of the X-IDS ecosystem. After formalizing the definition of `explainability’ for X-IDS, we need to create explanations tailored to the stakeholders. Figure \ref{fig: user_tax} demonstrates a simple user and explanation taxonomy. We consider three major stakeholders based on their roles in this taxonomy including \textit{developers}, \textit{defense practitioners}, and \textit{investors}. Each of the stakeholder categories necessitates a different degree of explanation and visualization. Programmers and CSoC members are more familiar with the field and may want more complex explanations. Investors, on the other hand, may be more satisfied with summarized visualizations. Each user group performs varying tasks based on the explanations. Programmers will work to debug and increase the efficacy of the AI model. CSoC members will be tasked with protecting investor assets. Indirectly, investors will need to make hiring or budgeting decisions.
Future research is needed to determine the best types of explanations for each user group.

\subsection{Evaluation Metrics} \label{recommendation_evaluation_metrics}

\begin{figure}[htbp]
    \centering
    \includegraphics[scale=0.50]{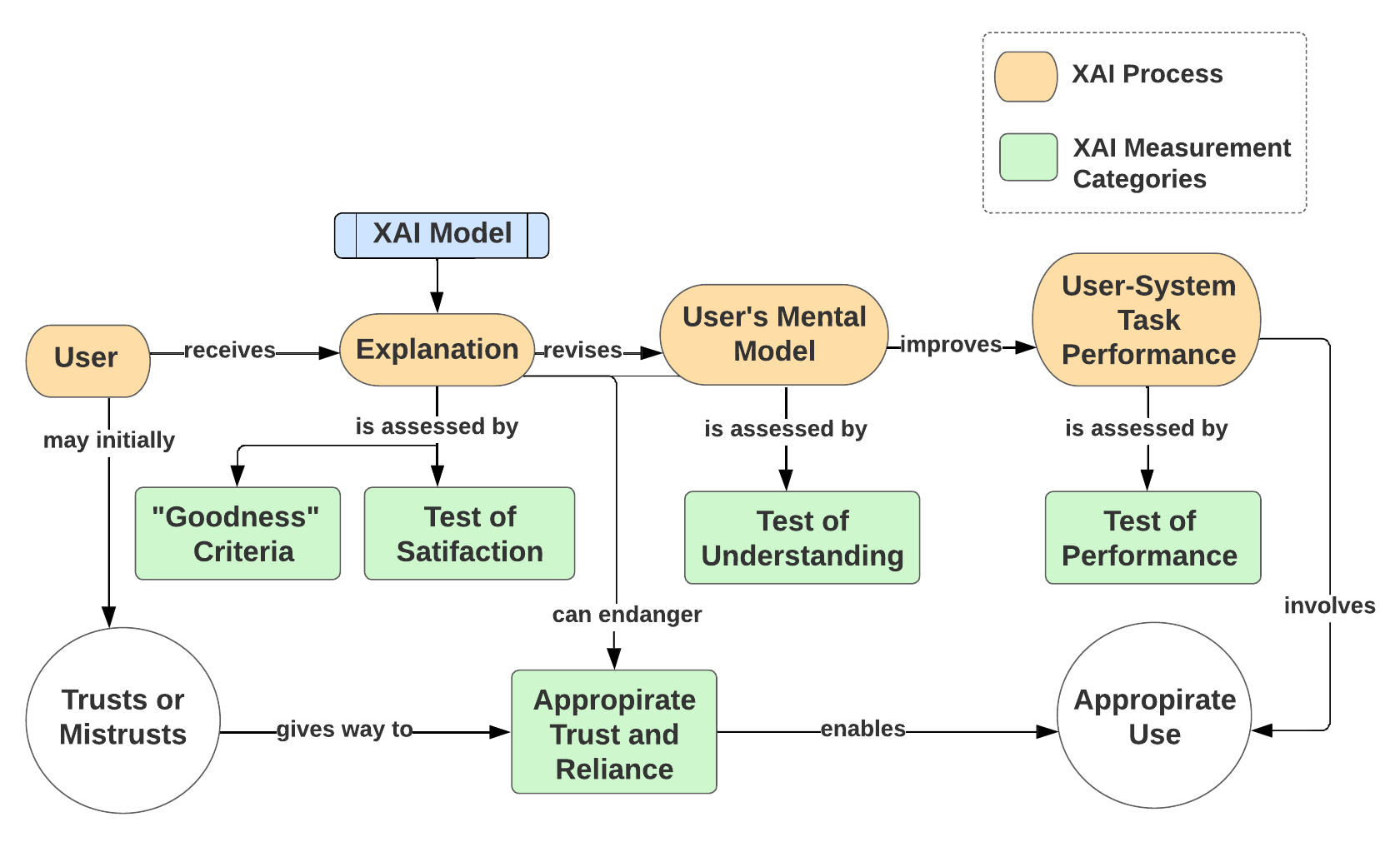}
    \caption{Different categories for assessing the effectiveness of explanations in the IHCM psychological model with detailed explanation process \cite{gunning2019darpa}.}
        \label{fig: explainability_metrics}
\end{figure}

The third challenge in designing X-IDS is evaluating the explanation generated by the `explainer module’. Finding the best explanation for each stakeholder category requires customized evaluation metrics. Currently, there is no consensus on metrics for explanations. In Section \ref{exp_metric} we described a body of literature proposing various evaluation metrics that could be used towards evaluating explanations. In particular, we recommended evaluation metrics proposed by authors in \cite{doshi2017towards} to evaluate explanations for X-IDS in Section \ref{exp_module_evaluation}. Another notable work that could serve as a baseline for evaluating explanations is the psychological model of explanation created by the Florida Institute for Human and Machine Cognition (IHMC) \cite{gunning2019darpa}. The proposed model is illustrated in Figure \ref{fig: explainability_metrics}. The user receives an explanation from the XAI model. This explanation can be tested for ``goodness'' and the satisfaction of the user/stakeholder. The user then revises their mental model of the XAI system. Their understanding of the system can be tested. Tasks are performed based on the explanation. The IHMC model merges the purpose of the XAI model, with the task and mindset of the user.A noteworthy method to evaluate the effectiveness of explanations is proposed by authors in \cite{gunning2019darpa}. Figure \ref{fig: explainability_metrics}  illustrates their approach.

\subsection{Adverserial AI}
Adversarial AI refers to the use of artificial intelligence for malicious purposes, including attacks on other artificial intelligence systems to evade detection \cite{anderson2016deepdga} or to poison data \cite{chen2017targeted}. Malicious actors can potentially attack the classifiers that are used to generate predictions and cause misclassification.  In context of X-IDS, the explanations generated by the explainer module may become a new point of attack for malicious actors. Attackers may  add, delete, or modify explanations to evade detection \cite{rastogi2022explaining}. Attackers may also attack training datasets to alter the explainer's behavior. The methods and effects of these attacks will need to be explored. Defense techniques must be created to correct attacked explanations. Studies to defend IDS against adversarial attack include \cite{han2021evaluating, pawlicki2020defending, hartl2020explainability}, etc. Study-specific to adversarial approach for X-IDS is discussed in \cite{marino2018adversarial}.

\subsection{Misleading/Incorrect Explanations}
An explanation does not have to be attacked to be misleading. The explanation itself may be misleading, or the user may interpret the explanation incorrectly. This may lead to circumstances where the model is correct and the user is the problem. The explainer will need to be modified to prevent user error in such situations.

Explanations that are misclassified either by an attack or due to the poor quality of data can have a significant negative impact on CSoCs. CSoCs security analysts should always critically analyze the reasoning behind the prediction. Moreover, methods for auditing previously incorrect explanations should be created. Ideally, the X-IDS should be able to audit itself and generate explanations for the audit. 

\subsection{Scalability and Performance}
Performance is of utmost importance for an IDS. 
CSoCs can incur losses for lost time. Explanations should not needlessly slow down an IDS. So how do we optimize an X-IDS? One approach is that the explainer could generate explanations for every sample it sees, or it could strategically choose which samples to explain. A comprehensive analysis of the CPU, RAM, and disk usage should be run on current and future explainers.



\section{Conclusion} 
\label{conclusion}

The exponential growth of cyber networks and the myriad applications that run on them have made CSoC, Cyber-physical systems, and critical infrastructure vulnerable to cyber-attacks. 
Securing these domains and their resources through the use of defense tools such as IDS, is critical to combating and resolving this issue \cite{wali2021explainable, wang2020explainable}. Recent AI-based IDS research has demonstrated unprecedented prediction accuracy, which is helping to lead to its widespread adoption across the industry. CSoC analysts largely rely on the results of these models to make their decision. However, in most cases, decision-making is impaired simply because these opaque models fail to justify their predicted outcomes. A solution to this problem is to embrace the concept of `explainability’ in these models. This, in turn, may facilitate quick interpretation of prediction, making it more feasible for CSoC analysts to accelerate response times.

A systematic review of current state-of-the-art research on `XAI’ or `explainability’ highlighted some key challenges in this domain, such as the lack of consensus surrounding the definition of `explainability’, the need to  formalize explainability from the user’s perspective, and the lack of metrics to evaluate explanations. We propose a taxonomy to address this problem with a focus on its relevance and applicability to the domain of intrusion detection.

In this paper, we present in detail two distinct approaches found in the body of literature which address the concern of `explainability’ in the IDS domain, including the white box approach and the black box approach. The white box approach makes the model in use inherently interpretable, whereas the black box approach requires post-hoc explanation techniques to make the predictions more interpretable (e.g., LIME \cite{ribeiro2016should}, SHAP \cite{lundberg2017unified}).While the former approach may provide a more detailed explanation to assist CSoC members in decision-making, its prediction performance is in general outperformed by the latter. Nevertheless, the field of IDS requires a high degree of precision to prevent attacks and avoid false positives. Bearing this in mind, a black box approach is recommended when developing an X-IDS solution. 

In addition, we also propose a three-layered architecture for the design of an X-IDS based on the DARPA recommended architecture \cite{gunning2019darpa} for the design of XAI systems. This architecture is sufficiently generic to support a wide variety of scenarios and applications without being bound by a particular specification or technological solution. Finally, we provide research recommendations to researchers that are interested in developing X-IDS.

\section{Acknowledgment}
This work by Mississippi State University was financially supported by the U.S. Department of Defense (DoD) High Performance Computing Modernization Program, through the US Army Engineering Research and Develop Center (ERDC) (\#W912HZ-21-C0058). The views and conclusions contained herein are those of the authors and should not be interpreted as necessarily representing the official policies or endorsements, either expressed or implied, of the U.S. Army ERDC or the U.S. DoD.

\bibliographystyle{unsrt}
\bibliography{refs,mittal}

\EOD

\end{document}